\documentclass[lettersize,journal]{IEEEtran}

\usepackage{amsmath,amsfonts}
\usepackage{algorithmic}
\usepackage{algorithm}
\usepackage{array}
\usepackage[caption=false,font=normalsize,labelfont=bf,textfont=normalfont]{subfig}
\usepackage{textcomp}
\usepackage{stfloats}
\usepackage{url}
\usepackage{xcolor}
\usepackage{verbatim}
\usepackage{graphicx}
\usepackage{cite}
\usepackage{multirow}

\definecolor{myblue}{RGB}{0,0,205}
\definecolor{myred}{RGB}{184, 115, 51}

\begin{document}

\title{Reformulating Parallel-Connected Lithium-Ion Battery Pack Dynamics with Interconnection Resistances as Ordinary Differential Equations}

\author{Jaffar Ali Lone, Nilsu Atlan, Simone Fasolato, Davide M Raimondo and Ross Drummond % <-this % stops a space
	\thanks{Jaffar Ali Lone thanks the Ministry of Education, Govt. of India, for providing financial support via PMRF Grant no. 2701805.}% <-this % stops a space
	%\thanks{Jaffar Ali Lone thanks the Ministry of Education, Govt. of India, for providing financial support via PMRF Grant no. 2701805.}
	\thanks{Jaffar Ali Lone is with the Department of Electrical Engineering, Indian Institute of Technology Patna, India. {\tt\small jaffar$\_$2121ee06@iitp.ac.in}}%
	\thanks{Nilsu Atlan and Ross Drummond are with the School of Electrical and Electronic Engineering, University of Sheffield, UK.\\ {\tt\small ross.drummond@sheffield.ac.uk}}%
       \thanks{Simone Fasolato is with the Department of Electrical, Computer and Biomedical Engineering University of Pavia, IT.\\ {\tt\small simone.fasolato01@universitadipavia.it}}
    \thanks{Davide Raimondo is with the Department of Engineering and Architecture, University of Trieste, Trieste, IT.\\ {\tt\small davide.raimondo@unipv.it}}
	%\thanks{$^{*}$Corresponding author.}% 
}



\maketitle
\begin{abstract}
This work presents analytical solutions for the current distribution in lithium-ion battery packs composed of cells connected in parallel, explicitly accounting for the presence of interconnection resistances. These solutions enable the reformulation of the differential-algebraic equations describing the pack dynamics into a set of ordinary differential equations, thereby simplifying simulation and analysis. Conditions under which uniform current sharing across all cells occurs are also derived. The proposed formulation is validated against experimental data and confirms its ability to capture the key behaviours induced by interconnection resistances. These results can support the improved design and control of parallel-connected battery packs.
\end{abstract}


\section{Introduction}
\IEEEPARstart{T}{oday}, typical lithium-ion batteries have cell-level energy densities of roughly $\approx100-500$ Wh~kg$^{-1}$\cite{frith2023non}-- significantly lower than those demanded by intensive applications such as electric aircraft and grid storage. For those applications, large battery packs formed by connecting several thousand individual cells in series and parallel are needed. Examples can be found in electric vehicles (such as the 7,104 cell pack of the Tesla Model S) and grid storage (such as the 18,900 cells used in \cite{slide} and the 50,000 cells used in \cite{vikrant2024ageing}). The move towards large packs in recent years has shifted the focus of modeling and analysis away from lumping all individual cells together and treating them uniformly, towards recognizing the pack as a complex system in which each unique cell contributes to the overall dynamics. 
This perspective is essential because battery designs that perform well at the cell level may not necessarily retain those advantages at the pack-level.  This was highlighted by Frith et al. \cite{frith2023non} in their discussion of a  \textit{graphite-SiOx}$||$\textit{NCA}  cell with an energy density of $\approx~$250 Wh kg$^{-1}$ which dropped to $\approx~$140 Wh kg$^{-1}$ at the pack-level. In contrast, a \textit{graphite}$||$\textit{LFP} cell with a significantly lower energy density of $\approx~$180 Wh kg$^{-1}$  had a comparable pack-level energy density of $\approx~$120 Wh kg$^{-1}$. In other words, higher cell-level energy densities do not necessarily lead to substantial performance gains at the pack-level. 
  A range of factors contribute to this energy density gap, including the enhanced thermal and electrochemical stability of LFP cathodes which reduces the need for active cooling. As pack sizes increase, additional challenges emerge that contribute to reduced efficiency \cite{frith2023non} and accelerate degradation \cite{bhaskar2024heterogeneity,naylor2024degradation, he2024numerical}. These include state-of-charge \cite{zhang2021cell} and temperature \cite{naylor2024degradation,ruwald2024experimental} heterogeneities, as well as scalability issues in designing battery management system algorithms \cite{zhang2020state,lone2022functional}.   

 Mathematical modeling can be used to reduce the performance drop-off between lithium-ion cells and packs, by offering deeper insight into pack-level behavior. In recent years, cell-level lithium-ion battery modelling has advanced rapidly, with a wide variety of codes now available including the electrochemical models of \texttt{PyBaMM} \cite{sulzer2021python}, \texttt{LIONSIMBA} \cite{torchio2016lionsimba}, \texttt{TOOFAB} \cite{khalik2021model}, and \texttt{PETLION} \cite{berliner2021methods}. However, these benchmark codes currently do not scale effectively to the pack-level, especially in the case of parallel connections. Existing methods, such as \texttt{Liionpack} \cite{tranter2022liionpack}, have limitations including having to  numerically solve the differential algebraic equations of Kirchhoff's laws at every time instant; as opposed to simply propagating the solution of an ordinary differential equation forward in time. 
The popularity of these codes highlights the need to extend recent advances in cell-level modelling to the pack-level, and so accelerate the development of open-source pack-level codes with comparable user-friendliness and simulation speeds.% \color{red} I WOULD ADD THE FOLLOWING (SEE IF THIS IS THE RIGHT PLACE) An approach to improving the scalability of electrochemical model simulations for large battery packs was proposed in \cite{saccani2022computationally}, leveraging the Waveform Relaxation method to decompose the system into smaller, more efficient sub-problems, which are solved iteratively to ensure convergence. However, the method provides improved computation time only when a sufficient number of cells are connected in parallel.

\begin{figure*}[t]
    \centering
    \subfloat[\small A parallel-connected lithium-ion battery pack \textit{without} interconnection resistances.\label{fig:pack_rd}]{
        \includegraphics[width=0.48\textwidth]{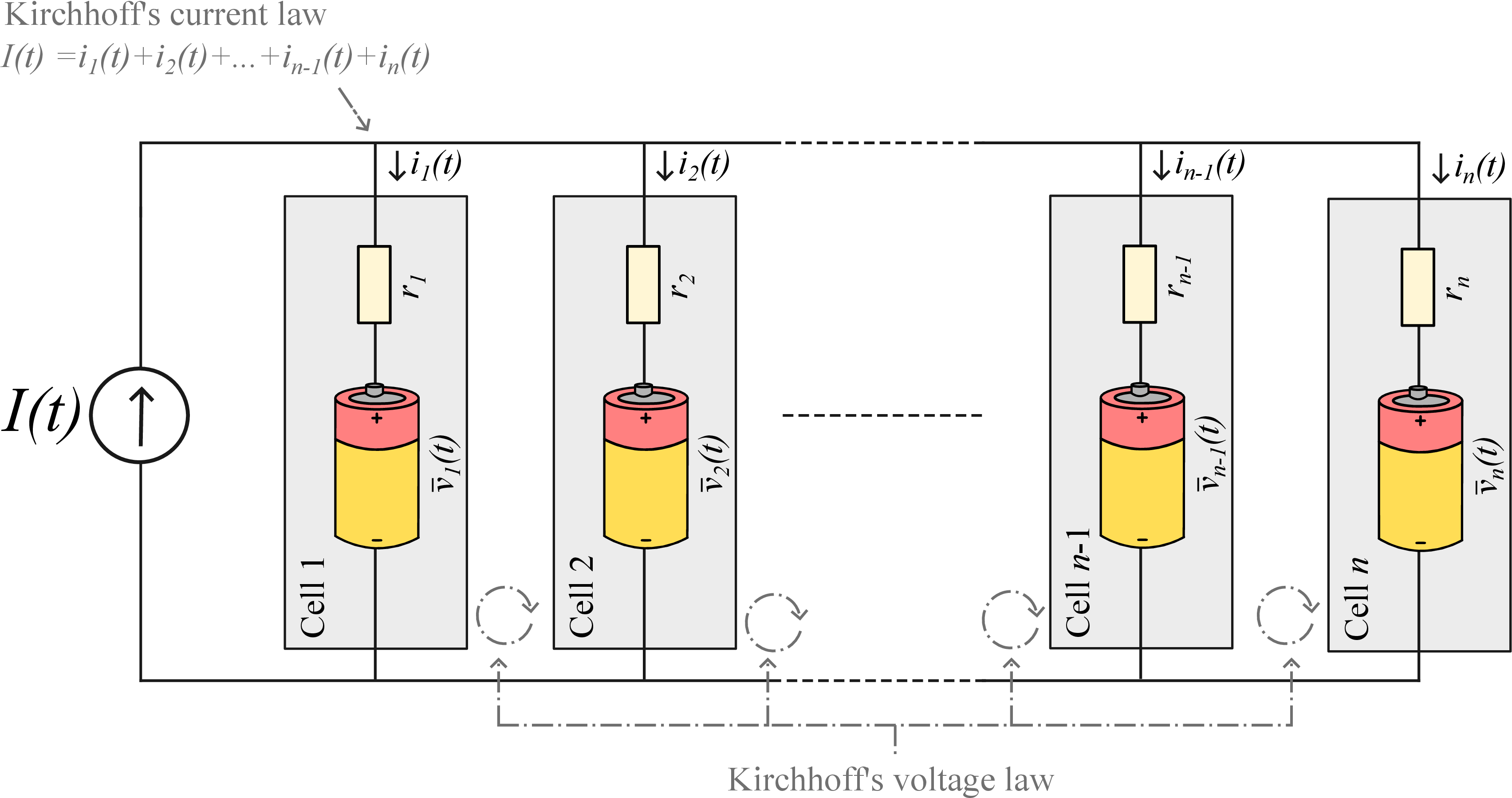}
    }
    \hfill
    \subfloat[\small A parallel-connected lithium-ion battery pack \textit{with} interconnection resistances $R_k$.\label{fig:pack_slide}]{
        \includegraphics[width=0.48\textwidth]{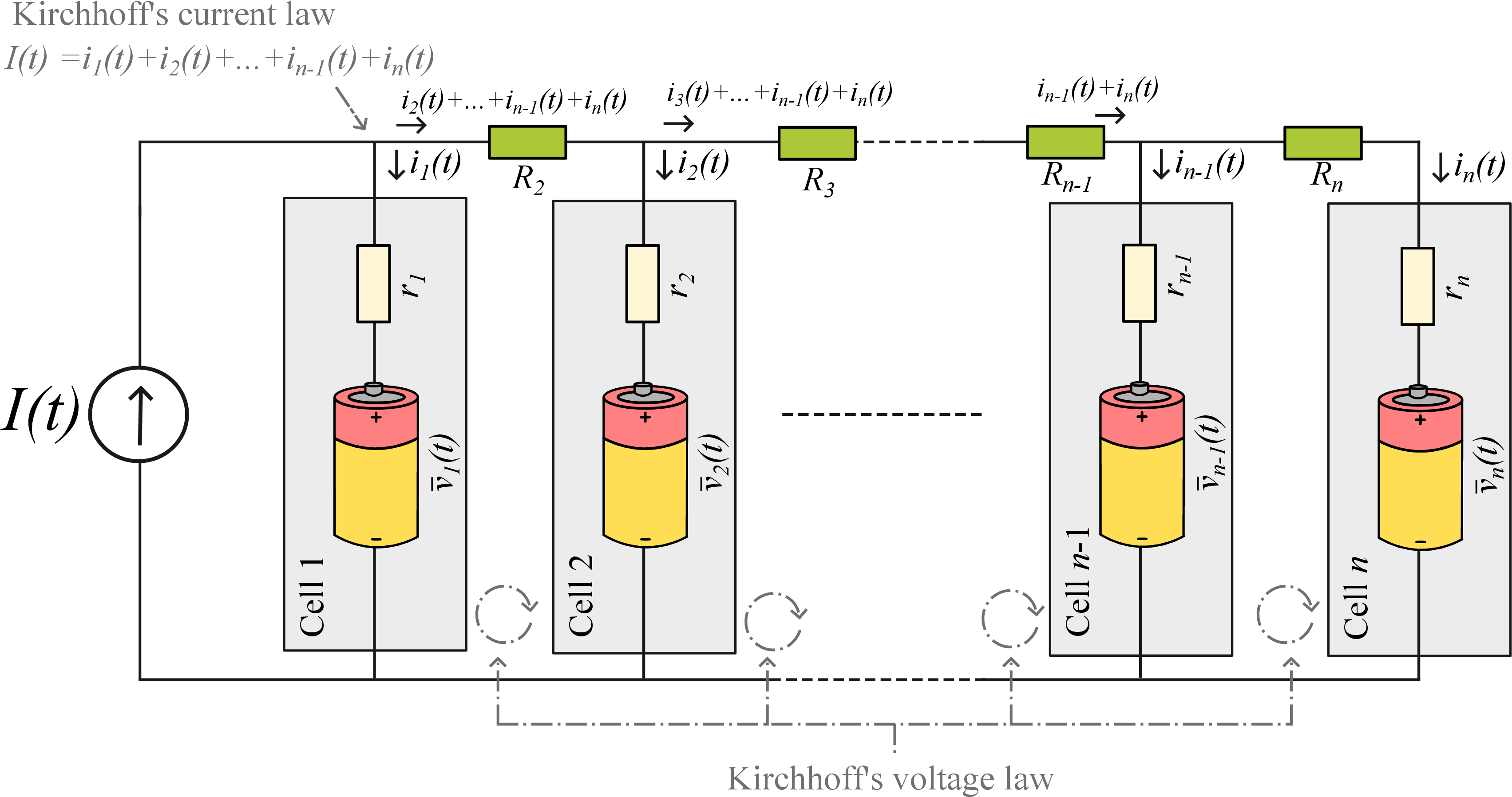}
    }
    \caption{Two parallel-connected battery packs: (a) ideal case without interconnection resistances $R_k$, and (b) with $R_k$ accounting for wire and interconnection resistances. The study examines how branch currents $i_k(t)$ distribute in response to these resistances.}
    \label{fig:packs}
\end{figure*}

%\color{black}
To accurately model battery packs, the effects of connecting cells in series and in parallel must be taken into account. Series connections increase the pack voltage and thus contribute to higher power output, while parallel connections increase the overall capacity, enhance fault tolerance to open-circuit failures, and may enable partial self-balancing under certain conditions. Modelling cells in series is relatively straightforward as each cell receives the same current, allowing for the decoupling of their dynamics (although cells in series can still be thermally coupled together). Modelling cells in parallel is more complex since algebraic equations must be solved at each time step to determine the branch current distribution (i.e., the current flowing into each parallel branch-- see Figure \ref{fig:packs} for an illustration) to ensure that Kirchhoff's current and voltage laws are satisfied (see Equations \eqref{A22_1} and \eqref{A22_2} and also \cite{dubarry2016cell, bruen2016modelling,hosseinzadeh2019combined}). These algebraic equations result in parallel pack models being described by \textit{differential-algebraic equations} (DAEs) rather than \textit{ordinary differential equations} (ODEs). %DAEs  typically require specialized numerical solvers with small step sizes to maintain solution accuracy and stability \cite{petzold1982differential}.  As a result, DAE models can be highly nonlinear and produce complex solutions; for parallel packs, they can result in non-trivial behaviours such as branch current fluctuations (see Figs. \eqref{BRC_LFP_RZ} and \eqref{BRC_LFP_RNZ}). %This added complexity makes DAE models computationally demanding and more difficult to implement, especially in real-time battery management systems. %To simplify this sitution, many modeling approaches first compute the numerical solutions to Kirchhoff’s laws, projecting the state of the index-1 DAE down into an ODE framework \cite{bruen2016modelling, chang2019correlations}. This approach reduces the complexity while still capturing the essential dynamics of the parallel pack, but it underscores the need for sophisticated computational methods when dealing with large-scale, parallel-connected battery systems.

 %Mathematical models are important tools for understanding  battery pack dynamics, and so how to optimise then. Battery packs are combinations of cells in series and parallel, with the series parts providing power and the parallel parts providing capacitance, fault-tolerance, and natural cell balancing. When the current is considered the input to the battery pack, as is usually the case, modelling cells in series is often trivial; as every cell just experiences the same current and so responds independently. However, modeling cells in parallel is more challenging. This is because 

% Several models have been proposed for parallel lithium-ion battery packs. 
Solving DAEs is, in general, more computationally demanding than solving ODEs—a challenge that has motivated studies on tailored numerical methods \cite{petzold1982differential}. For example, in \cite{saccani2022computationally}, the authors proposed a method to improve the scalability of electrochemical model simulations for large battery packs composed of parallel-connected cells governed by DAEs. The approach leverages the Waveform Relaxation technique to decompose the system into smaller, more computationally efficient sub-problems, which are solved iteratively until convergence. While effective, the method yielded significant reductions in computation time only when a sufficiently large number of cells were connected in parallel.

The inherent complexity of DAE-based formulations has motivated the development of methods that first convert the DAEs into ODEs by solving the algebraic equations. For packs composed of parallel-connected cells, this requires solving Kirchhoff's laws to express the distribution of branch currents in terms of the model states and the applied current.
Kirchhoff's laws can be solved either numerically or analytically.
The \emph{numerical} approach typically involves inverting a matrix—referred to in this paper as the $A_{22}$ matrix, as defined in Eqn. \eqref{a22inv} and in works such as \cite{bruen2016modelling,hosseinzadeh2019combined, vikrant2024ageing}. Beyond the computational burden of performing matrix inversion at each time step, this approach often lacks the physical insight needed for pack-level design optimization. In contrast, \emph{analytical} solutions avoid repeated matrix inversion—which can scale in complexity as $\mathcal{O}(n^3)$ where $n$ is the matrix dimension \cite{golub2013matrix}—and reveal how branch currents redistribute throughout the pack. These insights can support better pack design decisions, reduce degradation gradients \cite{bhaskar2024heterogeneity,naylor2024degradation,weng2024current,zhang2025study}, minimize current and parameter heterogeneity  \cite{ross2024comparison, telmasre2023initial,preger2025impact,ayalasomayajula2025physics}, and improve algorithms for pack-level battery management systems \cite{zhang2021cell}.  
 \color{black} 

 Existing analytical solutions for current distributions in battery pack models include those presented in \cite{drummond2021resolving,lee2025state}, as well as earlier studies such as \cite{fill2018current},  which focused on cell models with linear open-circuit voltage curves, and \cite{hofmann2018dynamics}, which neglected state-of-charge dynamics.  A key limitation of these prior solutions is that they did  not account for the interconnection resistances between cells, such as those arising from bus bars or wiring. In many practical applications, interconnection resistances can exceed the internal resistances of the cells themselves, leading to significant heterogeneity across the pack. This point was emphasized by Piombo   et al. \cite{piombo2024unveiling}, who stated that: \begin{quote}``\textit{According to the results, the interconnection resistance is the most relevant contributor to heterogeneous performance within the (parallel) string.}'' 
 \end{quote}
 However, existing analytical modelling approaches have not been generalized to battery packs with interconnection resistances. 
 %Instead,  numerical methods have been adopted including the scheme of \cite{slide} where the problem of enforcing the voltage constraints across the pack was interpreted as a control problem.
{Instead, alternative methods, such as the one proposed by \cite{slide}, have been adopted. In that paper, rather than directly enforcing the algebraic equations of Kirchhoff's laws in the parallel pack model equations,  a control theory approach was adopted with a proportional-integral  controller regulating the currents amongst the parallel-connected cells (based on voltage differentials relative to the mean voltage of all cells) to drive the error in the voltage constraints to zero. Although this approach enables dynamic current balancing, it lacks closed-form expressions for the branch currents and relies on appropriate tuning of control parameters to ensure stability. }

 This paper addresses this gap by generalizing explicit solutions for branch currents to account for interconnection resistances in the pack (see Section \ref{sec:slide}).
\\
 \textbf{Contributions:} The main contributions of this paper are:
 \begin{enumerate}
     \item Derivation of analytical  expressions for branch current distributions in battery packs composed of parallel-connected cells, as functions of the model states and the applied current, both with and without interconnection resistances.

     \item Application of these analytical  expressions to transform the DAE-based pack model into an ODE formulation. This reformulation offers computational advantages and provides deeper insight into current distributions across the pack. %Applying these explicit solutions to express DAE-based pack models as ODEs. \color{red} and show the computational and other (standart observers, insight on current distributions, etc.) advantages of this approach \color{black}

    \item Experimental validation of the proposed analytical solution using measured data. Simulations also illustrate the emergence of current fluctuations  within the pack. % \color{red} Heterogeneity in terms of what? Current is mentioned in the fluctuations. Cells are assumed to be identical in the simulations, right? And temperature is not accounted for. Is it in terms of SOC? Maybe better to clarify\color{black} 

\item Identification of pack configurations with interconnection resistances that enable uniform cell responses. In particular, generalized QR-matching conditions are derived for packs with interconnection resistances.
          
 \end{enumerate}
To the best of the authors’ knowledge, this is the first derivation of analytical current distributions for parallel-connected battery packs that account for interconnection resistances. %This approach not only has the potential to accelerate computation times by eliminating the need for matrix inversion (see Sections \ref{sec:A22inv} and \ref{sec:reccurence}), but also offers  insights into improving pack-level design and developing fault detection algorithms. 
 The theoretical results are validated against experimental  data and simulations of packs considering both LiNi$_x$Mn$_y$Co$_{1-x-y}$O$_2$ (NMC) and LiFePO$_4$ (LFP) cells. Code to run the models used in this paper can be found at \texttt{\url{https://github.com/jaffarlone07/Parallel-Pack-Paper-MATLAB-Code.git}}.

\section{Modeling Packs of Parallel-Connected Cells} \label{Modeling_sec}

We consider the problem of modeling battery packs composed of parallel-connected cells, as illustrated in Figures~\ref{fig:pack_rd} and ~\ref{fig:pack_slide}. For simplicity, we use the term \emph{pack} to refer to any collection of connected cells. This simplified use of the term pack is intended to streamline the language of the paper, although we acknowledge that, in practice, packs are often assembled from smaller groupings of cells—such as modules—and this terminology may introduce some ambiguity.

\subsection{Dynamics of a single cell}

The dynamics of each cell in the pack, denoted as $h_k(x_k(t),i_k(t)): \mathbb{R}^m \times \mathbb{R} \to \mathbb{R}^m$  are described by
\begin{subequations}\label{cell_mod}\begin{align}
\frac{d}{dt}x_k(t) = h_k(x_k(t),i_k(t)), \quad \forall k = 1,\,2,\,\dots,\,n,
\end{align}
where $x_k\in \mathbb{R}^m$ is the state of the $k^\text{th}$ cell in the parallel pack and $i_k(t) \in \mathbb{R}$ denotes the branch current it receives. The voltage of cell $k$, $v_k(t) \in \mathbb{R}$, is assumed to include a  resistance term
\begin{align}
    v_k(t) = \bar{v}(x_k(t)) + r_ki_k(t) \quad \forall k = 1,\,2,\,\dots,\,n.
\end{align}\end{subequations}
Here, $\bar{v}(x_k(t))$ represents the state-dependent component of cell $k$'s voltage, incorporating effects such as the open-circuit voltage, while $r_k$ represents its series resistance. As illustrated in Figure \ref{fig:packs}, the individual cell models are interconnected, with $R_k$ representing the interconnection resistance between them.  A key assumption in this work is that the Ohmic drop,  $r_ki_k(t)$,  is a linear function of the current. This linearity is fundamental, as it enables the branch currents  $i_k(t)$ to be computed using linear algebra.  While some battery models include nonlinear resistances, such as the  $\sinh^{-1}$ term arising from the homogenization of Butler-Volmer kinetics in single-particle models \cite{moura2016battery}, many widely used models assume linear resistances, including most equivalent circuit models \cite{nejad2016systematic} and the Doyle-Fuller-Newman model \cite{doyle1993modeling,drummond2019feedback}. Furthermore, because current fluctuations in uniform packs are typically small  \cite{luca2021current}  and the dynamics remain stable \cite{li2022demonstrating},  linearising the nonlinear resistances often yields sufficiently accurate approximations.  In contrast, the nonlinearity of the state-dependent voltage component $\bar{v}(x_k(t))$-- such as the open-circuit voltage-- does not impact the analysis.

The cell-level model structure in Eqn.  \eqref{cell_mod}  is intentionally kept general to emphasize the broad applicability of the analytical solutions for branch currents $i_k(t)$ developed in Sections \ref{sec:A22inv} and \ref{sec:reccurence}. This modeling framework encompasses a wide range of battery models, including equivalent circuit models with thermal dynamics, as well as certain classes of electrochemical models, such as the Doyle–Fuller–Newman electrochemical model \cite{drummond2019feedback,drummond2021resolving,doyle1993modeling,lambert2023detecting}. 

\subsection{Dynamics of battery packs with parallel-connected cells}
This paper investigates the behaviour of   $n$  parallel-connected cell models  from Eqn. \eqref{cell_mod}, as illustrated in Fig. \ref{fig:packs}. Two parallel  configurations are considered:
\begin{itemize}
    \item One that includes interconnection resistances (Section \ref{sec:slide} and Figure \ref{fig:pack_slide}).
    \item  Another that neglects them (Section \ref{Resolving} and Figure \ref{fig:pack_rd}). 
\end{itemize}
The configuration including interconnection resistances (Fig. \ref{fig:pack_slide})  follows the setup in \cite{slide}, while the other without resistances (Fig. \ref{fig:pack_rd}) mirrors \cite{drummond2021resolving}. Specifically,  \cite{drummond2021resolving} models $n$ parallel-connected cells without interconnection resistances (i.e., resistances between cell bus bars), whereas  \cite{slide} accounts for wiring resistances commonly found in practical setups \cite{luca2021current,slide}. These resistances, denoted as $R_k$, play a crucial role in shaping current distributions \cite{piombo2024unveiling}, as they reduce the amount of current reaching cells located further along the electrical path within the pack. %as they can lead to uneven current distributions among the cells, thereby providing a more accurate representation of the battery pack's real-world operational conditions. 

For both pack configurations, the objective is to derive analytical expressions for the current distribution $i_k(t)$ flowing into each cell in terms of the model states and the applied current $I(t)$. To proceed, we define the pack's state vector and the vector of branch currents
$i_k$  entering each cell as follows
$$
x(t) = \begin{bmatrix}x_1(t) \\ x_2(t) \\ \vdots \\ x_n(t) \end{bmatrix},
\quad 
i(t) = \begin{bmatrix}i_1(t) \\ i_2(t) \\ \vdots \\ i_n(t) \end{bmatrix}.
$$
The vector of cell currents $i(t)$ arises from the distribution of the applied current  $I(t)$  among the parallel-connected cells, in accordance with Kirchhoff’s current law.%The goal of this work is to derive analytical solutions for this current distribution $i(t)$.  %Since we need to compute the branch currents $i_k(t)$ at each time step to resolve Kirchhoff's current and voltage laws within the pack (see Section \ref{sec:drum}), the parallel pack model is a \emph{differential-algebraic} system; it combines the dynamics of the cells  (e.g. Eqns. \eqref{cell_mod}) with the algebraic equations of Kirchhoff's laws, which are used to resolve the branch currents  \cite{lone2022functional}. The goal is to \emph{derive an analytical solution for the algebraic equations governing these currents} $i_k(t)$ and transform the pack dynamics into an ordinary differential equation (ODE) system.  \color{red}Isn't this part a bit repetitive? I have the feeling some of the concepts have been previously expressed. Maybe have a look\color{black}

% In order to charge the battery pack,  the branch currents $i_k(t)$ need to be calculated from the models BMS's known variables, as in the model's states $x(t)$ and the applied current $I(t)$. In other words, we want to get an equation for 
% \begin{align}
% i(t) = f(x(t),I(t)),
% \end{align}
% for some function $f(.,.)$ which we have to find, as we can then put this current into the dynamics of each individual cell, following the dynamics of Eqn. \eqref{cell_dyns}.

\section{Parallel Cells \textit{Without} Interconnection Resistances } \label{Resolving}

 Consider a battery pack composed of parallel-connected cells with no interconnection resistances. In this idealized scenario, the cells are assumed to be perfectly connected to bus bars, which are modeled as having zero resistance. This setup, illustrated in Fig.  \ref{fig:pack_rd}, has been extensively studied in previous works, such as \cite{drummond2021resolving} and later in \cite{lee2025state}, but is included here to complete the analysis and highlight the difficulty in generalising to packs with interconnection resistances. %However, the assumption of defect-free connections between the cells is often unrealistic in practice, as evidenced by the step differences between the branch currents in experimental data such as \cite{luca2021current}. \color{red}Again a bit redundant?\color{black}

\subsection{Kirchhoff's laws  with $R_k = 0$}\label{sec:drum}
 For the parallel pack of Fig.  \ref{fig:pack_rd} without interconnection resistances, Kirchhoff’s voltage law requires that the voltages of all individual cells in the pack are equal, i.e.
\begin{subequations}\label{Kirchhoff_e}
\begin{align}\label{KirchoffV_e}
v_j(t) = v_k(t), \quad \text{for all }k,j = 1,\dots n,
\end{align} \end{subequations}
or, alternatively,
\begin{subequations}\label{F_Kirchhoff}
\begin{align}\label{V_Kirchoff}
\bar{v}_j(x_j(t))+ r_{j}i_{j}(t)=& \bar{v}_k(x_k(t)) + r_k i_k(t), \nonumber\\
&\text{for all }k,j = 1,\dots n.
\end{align} \end{subequations}
Since  Eqn.  \eqref{KirchoffV_e} holds for any  $j$  if it holds for one, we can set  $j = 1$ in Eqn. \eqref{Kirchhoff_e} and express it in a more compact form
\begin{subequations}\label{Kirchhoff_F}\begin{align}\label{Kirchoff_V}
\bar{v}_1(x_1(t))+ r_{1}i_1(t)= \bar{v}_k(x_k(t)) + r_k i_k(t),~ \text{for all }k = 2,\dots n.
\end{align}
At the same time, Kirchhoff's current law requires that the branch currents, $i_k(t)$, sum to the applied current, $I(t)$, such that
\begin{align}\label{kcl}
\sum^n_{k = 1}i_k(t) = I(t).
\end{align}\end{subequations}

The algebraic equations defining Kirchhoff’s current law for the parallel pack in Fig. \ref{fig:pack_rd} can then be written in the matrix form
\begin{align}\label{A221}
A_{22}i(t) = q(x(t)),
\end{align}
with
\small \begin{align}
q(x(t)) = \begin{bmatrix}I(t) \\ \bar{v}_2(t)-\bar{v}_1(t) \\ \bar{v}_3(t)-\bar{v}_1(t)\\ \vdots \\ \bar{v}_n(t)-\bar{v}_{1}(t)
 \end{bmatrix},~ 
A_{22}= \begin{bmatrix} 1 & 1 & \dots  & \dots & 1  \\ r_1 & -r_2 & 0  & \dots & 0 \\ r_1& 0 & -r_3 & \ddots & \vdots
\\
\vdots  & \vdots & \ddots & \ddots & 0
\\ r_1 & 0 & \dots & 0 & -r_n \end{bmatrix}.\label{A22_1}
\end{align}
 The branch currents could then be determined by inverting the $A_{22}$ matrix
\begin{align} \label{a22inv}
i(t) = {A_{22}}^{-1}q(x(t)).
\end{align}
 By examining the structure of matrix $A_{22}$, and noting that $r_k~\geq~0,~\forall k= 1,\cdots,n$, it follows that its columns are linearly independent either if $r_1>0$ or, in the case $r_1=0$, if and only if $r_k>0, \text{for all } k=2,\cdots,n$. The following section shows how these branch currents could instead be computed analytically, without having to invert the $A_{22}$ matrix, by simply solving Eqn. \eqref{A221}.

\color{black} \subsection{Distribution of branch currents with $R_k = 0$}\label{sec:A22inv}

%To derive the analytical expression for the branch currents  $i_k$ in terms of the pack's states $x(t)$ and the applied current $I(t)$, Kirchhoff's laws, as presented in \eqref{Kirchhoff_F}, are first expressed as follows:
Kirchhoff's voltage laws of Eqn. \eqref{Kirchhoff_F} can be expressed as
% \begin{subequations}\begin{align} 
% r_1i_1(t)-r_2i_2(t)  & = \Delta\bar{v}_{21}(t), 
% \\ 
% r_1i_1(t)-r_3i_3(t)  & = \Delta\bar{v}_{31}(t), 
% \\ \nonumber
% & \vdots \\ 
% r_1i_1(t)-r_ni_n(t)  & = \Delta\bar{v}_{n1}(t),
% \\
% i_1(t)+i_2(t)+ \cdots + i_n(t) &=I(t), \label{i_tot}
% \end{align}\end{subequations}
% .  With this formulation, each of the branch currents, $i_k(t)$, can be expressed in terms of the first one, $i_1(t)$, as in
\begin{subequations}\begin{align} 
i_2(t)& = \frac{1}{r_2}(r_1i_1(t)-\Delta\bar{v}_{21}(t)), 
\\ 
i_3(t)& = \frac{1}{r_3}(r_1i_1(t)-\Delta\bar{v}_{31}(t)), 
\\ \nonumber
& ~\, \vdots  \\
i_n(t)& = \frac{1}{r_n}(r_1i_1(t)-\Delta\bar{v}_{n1}(t)),
\end{align}\end{subequations}
using the notation  $\Delta\bar{v}_{jk}(t)=\bar{v}_j(t)-\bar{v}_k(t)$. Substituting these expressions into the current law of \eqref{kcl} gives
 \begin{align}
     I(t) &=i_1(t)+i_2(t)+ \cdots + i_n(t), \nonumber\\
     &= i_1(t)+\sum^n_{\ell = 2}\frac{1}{r_\ell}(r_1i_1(t)-\Delta\bar{v}_{\ell1}(t)).
 \end{align}
The first branch current, $i_1(t)$,  can then be expressed in terms of the model’s state and the applied current $I(t)$ only
\begin{subequations}\label{eqn:ik1}\begin{equation} \label{first_br_cur}
    i_1(t)=\Biggr(r_1\sum^n_{\ell = 1}\frac{1}{r_\ell}\Biggr)^{-1}\Bigg(\sum^n_{k = 2}\frac{\Delta\bar{v}_{k1}(t)}{r_k}+I(t)\Bigg). 
\end{equation}
Using \eqref{first_br_cur}, the currents of the other cells can be expressed as
\footnotesize \begin{align}
    i_j(t)&=\frac{1}{r_j}\left(\Biggr(\sum^n_{\ell = 1}\frac{1}{r_\ell}\Biggr)^{-1}\Bigg(\sum^n_{k = 2}\frac{\Delta\bar{v}_{k1}(t)}{r_k}+I(t)\Bigg)-\Delta\bar{v}_{j1}(t)\right)\\ \nonumber
    &\text{for all }j = 2,3, \dots n.
\end{align}\end{subequations}
Therefore, in the absence of interconnection resistances, the currents of parallel-connected cells are defined by Eqn. \eqref{eqn:ik1}. This means that there is no need to invert the  $A_{22}$ matrix of Eqn. \eqref{A22_1}. Substituting Eqns. \eqref{eqn:ik1}  into the cell-level models in Eqn. \eqref{cell_mod} allows the parallel pack model to be reformulated as a set of ordinary differential equations, as the algebraic variables $i(t)$  become explicit functions of the differential states and the applied current.

\section{Parallel Cells \textit{With} Interconnection Resistances }\label{sec:slide}

In practical settings, particularly in prototype designs, battery packs exhibit interconnection resistances between cells. The algebraic expression for the branch currents given in Equation \eqref{first_br_cur} is no longer valid in the presence of these resistances, as the additional ohmic drops influence the current distribution within the pack. This section addresses the issue by deriving an expression for the current distribution in the configuration shown in Figure \ref{fig:pack_slide}, explicitly accounting for interconnection resistances.  
To the best of the authors' knowledge, analytical expressions for current distributions in parallel-connected packs with interconnection resistances have not been previously reported and represent a central contribution of this work. Using the derived currents   (see Eqn. \eqref{eqn:sn}), the DAE model of the parallel pack can be reformulated as an ODE involving only the model parameters, state variables, and the applied pack current.

\subsection{Kirchhoff's laws with $R_k \neq 0$}
Consider the pack shown in \cite[Fig. 4]{slide} and reproduced here in Fig. \ref{fig:pack_slide}. The difference between this pack and the one from \cite{drummond2021resolving} lies in the interconnection resistances $R_k$ between the cells.
In this case, Kirchhoff's current and voltage laws take the form %(as in the sum of voltages around each loop in Fig. \ref{fig:pack_slide}) 

\begin{subequations}\label{kv_1_new}\begin{align}
\sum^n_{k =1}i_k(t)  & = I(t),
\\
 r_1i_1(t)+\bar{v}_1(t)   = R_2\left(\sum_{k =2}^ni_k(t)\right)+ r_2&i_2(t)+\bar{v}_2(t),  \label{eqn_2}
\\
 r_2i_2(t)+\bar{v}_2(t)   = R_3\left(\sum_{k =3}^ni_k(t)\right)+ r_3&i_3(t)+\bar{v}_3(t), 
\\ \nonumber
  & ~\, \vdots \\
 r_{n-1}i_{n-1}(t)+\bar{v}_{n-1}(t)  = R_ni_n(t)&+ r_ni_n(t)+ \bar{v}_n(t). \label{eqn_end}
\end{align}\end{subequations}
This system of equations can be written in a form analogous to   Eqn. \eqref{A22_1} with
\begin{align}
A_{22}i(t) = q(x(t))
\end{align}
and
{\small \begin{align} \label{A22_2}
A_{22}= \begin{bmatrix} 1 & 1 & \dots  & \dots & 1  \\ r_1 & -(r_2+R_2) & -R_2  & \dots & -R_2 \\ 0 & r_2 & -(r_3+R_3) & \ddots & -R_3
\\
\vdots  & \ddots & \ddots & \ddots & R_{n-1}
\\ 0 & \dots & 0& r_{n-1} & -(R_n+r_n) \end{bmatrix}.
\end{align}}

\normalsize \subsection{Distribution of branch currents with $R_k \neq 0$}\label{sec:curr_R}

For the parallel pack shown in Fig.  \ref{fig:pack_slide}, the objective of this section is to express the pack's branch currents, $i_k(t)$, in terms of its states, $x(t)$, and the applied current, $I(t)$. To achieve this, Kirchhoff's voltage laws, as outlined in \eqref{eqn_2}-\eqref{eqn_end}, are first rewritten as
% \begin{subequations}\label{kvl_inter}\begin{align}
% r_{n-1}i_{n-1}(t)-(R_n+r_n)i_n(t)  & = \bar{v}_n(t)-\bar{v}_{n-1}(t),
% \\
% r_{n-2}i_{n-2}(t)-(R_{n-1}+r_{n-1})i_{n-1}(t)-R_{n-1}i_n  & = \bar{v}_{n-1}(t)-\bar{v}_{n-2}(t), \\ 
% \vdots \nonumber
% \\
% r_{1}i_{1}(t)-r_{2}i_{2}(t)-R_{2}\sum^n_{k =2}i_k  & = \bar{v}_2(t)-\bar{v}_{1}(t),
% \end{align}\end{subequations}
% which can be re-arranged to

\begin{subequations}\label{eqns_int}
\begin{align}
i_{n-1}(t)  &= \frac{1}{r_{n-1}}\left((R_n + r_n)i_n(t) + \bar{v}_n(t) - \bar{v}_{n-1}(t)\right), \label{ee_1}
\\
i_{n-2}(t)  &= \frac{1}{r_{n-2}}\Big((R_{n-1} + r_{n-1})i_{n-1}(t) + R_{n-1}i_n(t) \notag \\
&\hspace{3.5em} +\, \bar{v}_{n-1}(t) - \bar{v}_{n-2}(t)\Big), \label{ee_2}
\\
&~\, \vdots \notag
\\
i_{1}(t)  &= \frac{1}{r_1}\left(r_2 i_2(t) + R_2 \sum^n_{k=2} i_k(t) + \bar{v}_2(t) - \bar{v}_1(t)\right). \label{ee_3}
\end{align}
\end{subequations}

In contrast to Equation \eqref{first_br_cur},  expressing the first branch current $i_1(t)$  in terms of the model states,  via the voltages $\bar{v}_k(t)$ and the applied current $I(t)$, is no longer straightforward when interconnection resistances are present in the pack. To solve Eqn.  \eqref{eqns_int} for the branch currents $i_k(t)$, we first introduce the variables $S_k(t)$, defined as the sum of the currents from branch $k$ to $n$, as follows
\begin{align}\label{def_S}
    S_k(t) = \sum^n_{j =k}i_j(t).
\end{align}
Additionally, Kirchhoff's current law implies the following relations
\begin{align}\label{S_var}
    S_1(t) = \sum^n_{k =1}i_k(t) = I(t), \qquad S_n(t) = i_n(t)
\end{align}
which will be used later in the analysis (see Eqn. \eqref{eqsn1}).
%\subsection{Kirchhoff's laws in terms of $S_k$}\color{red}Do we need to start another subsection?\color{black}
Kirchhoff’s laws can then be expressed in terms of the  ``\textit{current sum}'' variables $S_k(t)$. Before proceeding, we define the following variables
$$\theta_k  = \frac{r_{k}}{r_{k-1}}, \quad \rho_k(t) = \frac{\bar{v}_{k}(t)-\bar{v}_{k-1}(t)}{r_{k-1}}, \quad
    \omega_k = \frac{R_k}{r_{k-1}}.$$
 Kirchhoff's voltage laws \eqref{eqns_int} can then be expressed as
\small \begin{subequations}\label{kvl1}
\begin{align}
i_{n-1}(t)  & = \theta_n i_n(t) +\omega_n i_n(t)+ \rho_n(t),
\\
i_{n-2}(t)  & = \theta_{n-1}i_{n-1}(t) +\omega_{n-1}(i_{n-1}(t)+i_n(t)) +\rho_{n-1}(t),
\\
i_{n-3}(t)  & = \theta_{n-2}i_{n-2}(t) +\omega_{n-2}\sum^n_{k = n-2}i_k(t) +\rho_{n-2}(t),\\
 & ~\, \vdots \nonumber
 \\
i_{2}(t)  & = \theta_{3}i_{3}(t) + \omega_3 \sum_{k =3}^ni_k(t)+\rho_3(t),
\\
i_{1}(t)  & = \theta_{2}i_{2}(t) + \omega_2 \sum_{k =2}^ni_k(t)+\rho_2(t).
\end{align}
\end{subequations}
\normalsize Using the fact that
\begin{align}
    i_k(t) = S_{k}(t)-S_{k+1}(t), \quad \forall k = 1,\dots, \, n-1,
\end{align}
then Eqns. \eqref{kvl1} can be expressed as
\begin{subequations}\label{S_diff}
\begin{align}
S_{n-1}(t) - S_n(t) &= \theta_n S_n(t) + \omega_n S_n(t) + \rho_n(t),
\\
S_{n-2}(t) - S_{n-1}(t) &= \theta_{n-1}(S_{n-1}(t) - S_n(t)) \notag \\
&\hspace{3em} +\, \omega_{n-1} S_{n-1}(t) + \rho_{n-1}(t),
\\
S_{n-3}(t) - S_{n-2}(t) &= \theta_{n-2}(S_{n-2}(t) - S_{n-1}(t)) \notag \\
&\hspace{3em} +\, \omega_{n-2} S_{n-2}(t) + \rho_{n-2}(t),
\\
&~\, \vdots \notag
\\
S_{2}(t) - S_{3}(t) &= \theta_3(S_3(t) - S_4(t)) + \omega_3 S_3(t) + \rho_3(t),
\\
S_{1}(t) - S_{2}(t) &= \theta_2(S_2(t) - S_3(t)) + \omega_2 S_2(t) + \rho_2(t).
\end{align}
\end{subequations}

After defining the parameters
$\alpha_{k}=  (1+\theta_{k}+~\omega_{k})$  for $ k~=~2,\dots, n, $ then Eqns. \eqref{S_diff} can be written more compactly as 
%Finally, rewriting Eqn. \eqref{S_diff} as
\begin{subequations} \label{fin_sn}
\begin{align}
S_{n-1}(t) & = \alpha_nS_n(t) + {\rho}_n(t),
\\
S_{n-2}(t)  & = \alpha_{n-1}S_{n-1}(t)-\theta_{n-1} S_{n}(t)+{\rho}_{n-1}(t),
\\
S_{n-3}(t) & =  \alpha_{n-2}S_{n-2}(t)-\theta_{n-2} S_{n-1}(t)+{\rho}_{n-2}(t),\\
& ~\, \vdots
 \\
S_{2}(t) & =  \alpha_{3}S_{3}(t)-\theta_{3} S_{4}(t)+{\rho}_3(t),
\\
S_{1}(t) & = \alpha_{2}S_{2}(t)-\theta_{2} S_{3}(t)+{\rho}_2(t).
\end{align}
\end{subequations}

\subsection{Recurrence formulas for the branch currents}\label{sec:reccurence}
Next, define two sequences $\beta_k$ and $f_k$ which propagate backwards through the pack from $k =n+1$ to $k = 2$.  Specifically, $\beta_k$ is defined by the recursion
\begin{subequations}\label{beta_seq}
\begin{align}
\beta_{n+1}  & = 1, \\
    \beta_n  & = \alpha_n,
    \\
    \beta_{k}  & = \alpha_{k}\beta_{k+1} - \theta_k\beta_{k+2}, \quad \forall k = 2,\,\dots, n-1,
\end{align} 
\end{subequations}
and $f_k$ by
\begin{subequations}\label{f_seq}
\begin{align}
f_{n+1}  & = 0,\\
    f_n  & = {\rho}_n(t),
    \\
    f_{k} & =\alpha_{k}f_{k+1} -\theta_{k}f_{k+2}+ {\rho}_{k}(t) \quad \forall k = 2,\,\dots, n-1.
\end{align} 
\end{subequations}

Using these two sequences,  Eqns.  \eqref{fin_sn} can express $S_{k-1}(t)$ in terms of $S_n(t)$ via
\begin{align}\label{Sk_relation}
S_{k-1}(t) & = \beta_k S_n(t) + f_k,  \quad k = 2,\,\dots,\, n.
\end{align}
This approach is similar to the one used in Eqn.  \eqref{first_br_cur} without interconnection resistances. However, when interconnection resistances are taken into account, the solution is expressed in terms of the cascaded current sums, $S_k(t)$. 
Using the recursive definitions of $f_k$ and $\beta_k$, as well as the relationship between Kirchhoff’s current law and $S_1(t)$ in \eqref{S_var}, it is now possible to write
\begin{align} \label{eqsn1}
    S_1(t) = \beta_2 S_n(t)+f_2 = I(t).
\end{align}
Solving Eqn. \eqref{eqsn1}  for  $S_n(t)$, and using the relation $S_n(t) = i_n(t)$ from \eqref{S_var}, gives
\begin{align}\label{eqn:sn}
   S_n(t)= \frac{I(t)-f_2}{\beta_2} = i_n(t).
\end{align}
The final branch current, $i_n(t)$, has now been expressed in terms of the model's states,  $x(t)$, and the applied current, $I(t)$.
Once $i_n(t)$ is determined, the remaining currents $i_{1:n-1}(t)$  can be obtained by backward propagation through the pack via \eqref{kvl1}. In other words, substituting Eqn. \eqref{eqn:sn} into \eqref{kvl1} provides an explicit solution for the branch currents without requiring the inversion of the  $A_{22}$ matrix in Eqn. \eqref{A22_2}. {This result  implies an inverse of $A_{22}$ exists as long as $\beta_2 \neq 0$, a condition defined by the sequence of Eqn. \eqref{beta_seq}}.

\begin{figure}[t]
    \centering
    \includegraphics[width=\columnwidth]{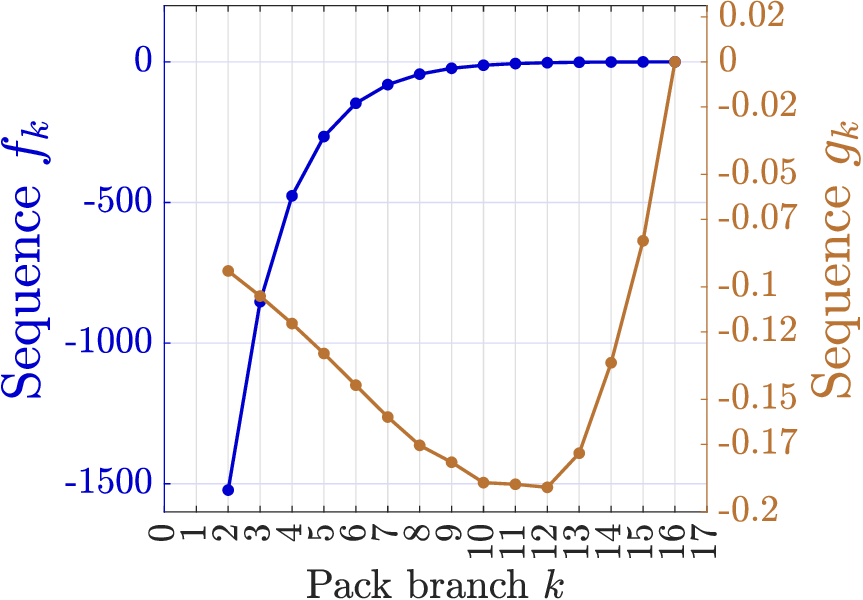}
\caption{\color{black} Sequences  $f_k$ (\color{myblue} navy blue\color{black}) and $g_k$ (\color{myred} orange\color{black}) at the end of a discharge profile for a parallel-connected pack of  $n = 15$ LFP cells with interconnection resistances and $c = 0.5$. The sequence $f_k$ grows exponentially as $k$ decreases from $n+1$ to $2$, while the scaled sequence  $g_k$ remains bounded, facilitating simulation and storage for large packs.}
\label{fig_fk}
\end{figure}

\begin{figure}[t]
    \centering
    \includegraphics[width=\columnwidth]{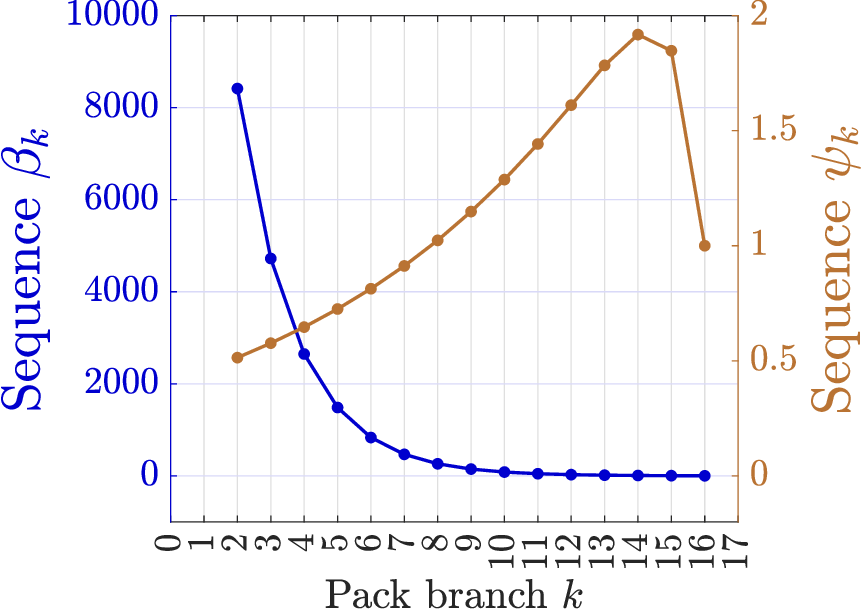}
\caption{\color{black} Sequences  $\beta_k$ (\color{myblue} navy blue\color{black}) and $\psi_k$ (\color{myred} orange\color{black}) at the end of a discharge profile for a parallel-connected pack of  $n = 15$ LFP cells with interconnection resistances and $c = 0.5$. Again, the sequences $\beta_k$ grows exponentially as $k$ decreases from $n+1$ to $2$, while the scaled sequence  $\psi_k$ remains bounded.}
\label{fig_betak}
\end{figure}

\subsection{Comments}
Section \ref{sec:curr_R} presented the derivation of the current distribution in the parallel pack of Fig. \ref{fig:pack_slide}, accounting for the interconnection resistances. Computing the branch currents requires first evaluating $i_n(t)$ using Eqn. \eqref{eqn:sn}, which is then used in Eqn. \eqref{kvl1} to determine the remaining currents $i_k(t)$. The evaluation of $i_n(t)$ itself involves a backward recursion through the pack to compute the sequences $\beta_k$ and $f_k$ defined in Eqns. \eqref{beta_seq}–\eqref{f_seq}. Numerical experiments suggest that the recursion in Eqns. \eqref{beta_seq}-\eqref{f_seq} can exhibit instability, potentially leading to numerical errors in large packs.
By first focusing on $f_k$, a rough explanation for the instability could be inferred by thinking of Eqn. \eqref{beta_seq} as a time-varying discrete-time system running backwards in time from $k = n+1$ to $k = 2$. The coefficients in this sequence satisfy $\alpha_k >1$ and $\alpha_k>\theta_k$, conditions that imply the recursion is unstable. This instability is evident in Figure \ref{fig_fk}, which shows exponential growth in the values of $f_k$, as $k$ propagates backwards, for an $n = 15$ LFP pack. %However, a more rigorous analysis is needed to bound the growth rate of these dynamics in future work.} 
Even though the dynamics of  \eqref{f_seq} are finite-time, the instability could introduce problems when storing the sequence $f_k$ when $n$ is large, as the values approaching $f_1$ would grow rapidly, slow down the computations, and reduce the solution accuracy. To avoid storing these large numbers, it is proposed to define a new variable 
\begin{align}
    g_k =c^{n+1-k}f_k,\label{gk}
\end{align}which dynamically scales $f_k$ by a scalar $c$. The dynamics for this scaled variable are
\begin{subequations}\label{g_seq}
\begin{align}
g_{n+1}  &= 0, \\
g_n      &= c\,\rho_n(t), \\
g_k      &= c\,\alpha_k\,g_{k+1} - c^2\theta_k\,g_{k+2} \notag \\
         &\hspace{3em} +\, c^{n+1-k}\,\rho_k(t), \quad \forall k = 2,\dots,n-1.
\end{align} 
\end{subequations}

Selecting $0 <c <1$ scales the dynamics, and, because all the dynamics are linear, it is then possible to recover $f_k$ from $g_k$ and so obtain the branch currents from \eqref{f_seq}.  Computing the branch currents in this way using the bounded $g_k$ sequence avoids the need to store the unstable variables $f_k$ and allows the method to scale to large packs. 

Figure \ref{fig_fk} demonstrates how the scaled sequence $g_k$ can reduce the effect of the instability in propagating through $f_k$. For this plot, $c = 0.5$ and the resulting $g_k$  is both stable and bounded, whereas $f_k$ grows exponentially as $k$ propagates backwards through the pack.

{The sequence $\beta_k$ suffers from similar instability issues as $f_k$. A similar fix to  Eqn. \eqref{gk}  can be found  by defining a new variable
\begin{align}\label{psi_k}
    \psi_k =c^{n+1-k}\beta_k
\end{align}
to transform the  $\beta_k$ dynamics from Eqn. \eqref{beta_seq} into
\begin{subequations}
\begin{align}
\psi_{n+1}  & = 1, \\
    \psi_n  & = c\,\alpha_n,
    \\
    \beta_{k}  & = c\,\alpha_{k}\psi_{k+1} - c^2\,\theta_k\psi_{k+2}, \quad \forall k = 2,\,\dots, n-1.
\end{align} 
\end{subequations}
The original sequence $\beta_k$ can then be obtained by inverting Eqn. \eqref{psi_k}. Figure \ref{fig_betak} compares  these two sequences for the pack setup of Figure \ref{fig_fk}. Again, it shows the instability of the original sequence, in this case $\beta_k$, and the convergent behaviour of the scaled one, $\psi_k$. As discussed above, this approach can enable the method to scale to large packs. }

%As far as the authors are aware, the solution  \eqref{eqn:sn} describes the first explicit solution for current distributions in parallel-connected packs with interconnection resistances. By doing so, it enables the conversion of the DAE pack model into an explicitly defined ODE model. Since ODE models are generally easier to analyse and simulate than DAEs, this formulation could, in the future, enhance pack-level control strategies, improve BMS algorithms, and deepen our understanding of cell-to-cell degradation heterogeneity.\color{red}Stopped here\color{black} 

\begin{figure*}[t]
    \centering
    \subfloat[\small Identical cells: darker lines correspond to higher-indexed cells.]{
        \includegraphics[width=0.31\textwidth]{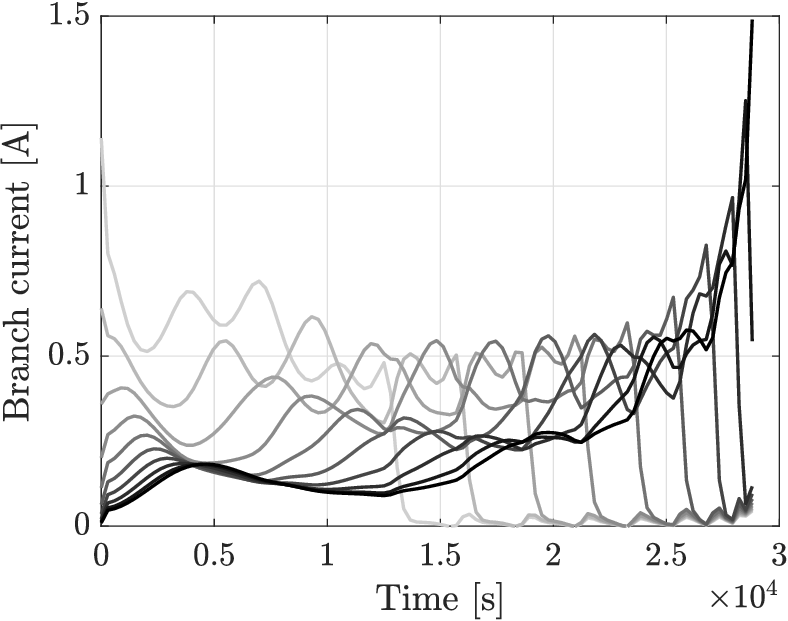}
        \label{fig:uneven}
    }
    \hfill
    \subfloat[\small Resistances tuned using \eqref{R_r}: uniform current sharing achieved.]{
        \includegraphics[width=0.31\textwidth]{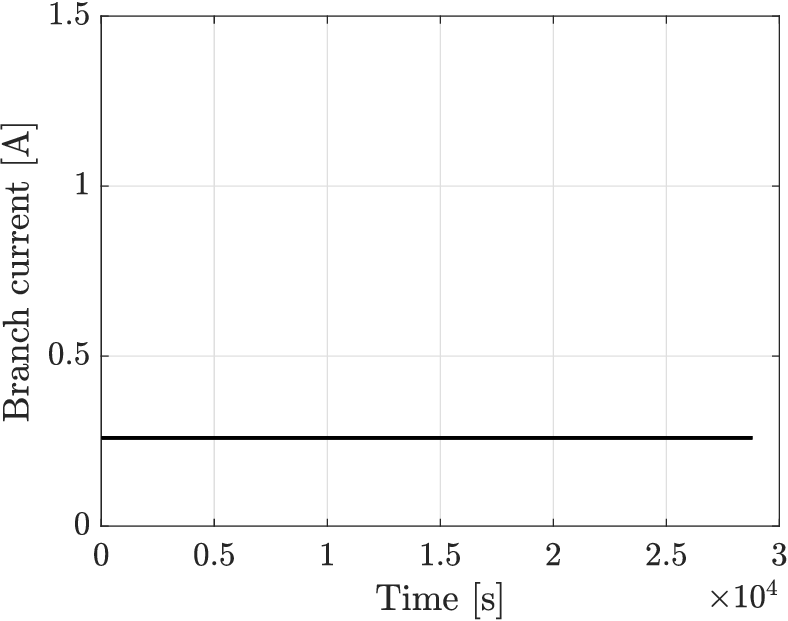}
        \label{fig:even}
    }
    \hfill
    \subfloat[\small Growth in current imbalance with increasing $R_k/r_k$.]{
        \includegraphics[width=0.31\textwidth]{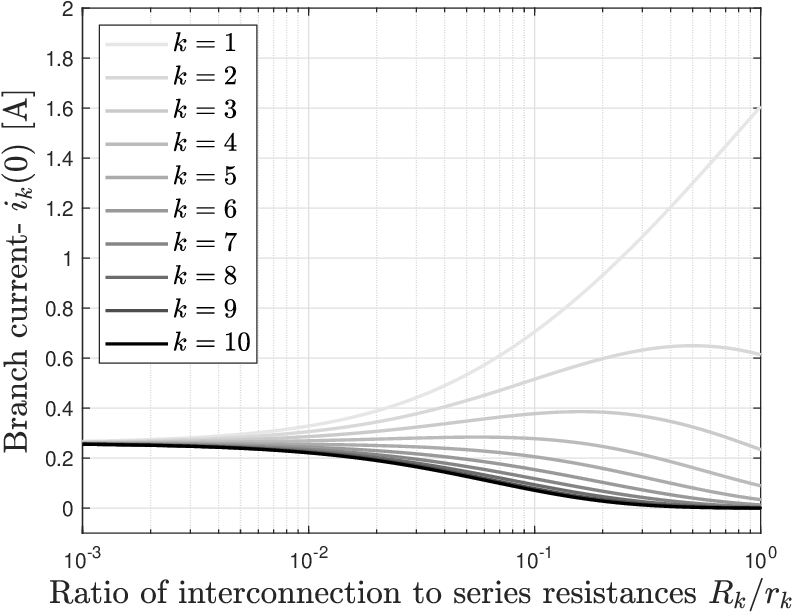}
        \label{eta}
    }
    \caption{Current distributions during 1C charging for ten LFP cells connected in parallel using the parameter values from Section~\ref{Simu_res}. Subfigure (a) shows the case with identical cells, resulting in uneven current distribution. Subfigure (b) demonstrates how tuning $R_k$ according to \eqref{R_r} leads to equal current sharing. Subfigure (c) illustrates the increase in current variation as the ratio $R_k/r_k$ increases.}
    \label{fig_all_even1}
\end{figure*}

\subsection{Uniform charging: Generalised QR-matching}

 \color{black} Whilst the focus of the above analysis has been on determining how current heterogeneities form across a pack, the analytic current distributions of Eqn.s \eqref{eqn:ik1} and \eqref{kvl1} can also be used to used to solve the reverse problem; how should the pack be designed such that all cells are used uniformly? For the case of two cells connected in parallel without interconnection resistances, this problem was solved by Weng et al. \cite{weng2024current} with their proposed QR-matching conditions. In particular, they considered cells with dynamics for the state-of-charge, $z_k(t)$,
\begin{align}
    \dot{z}_k(t) = \frac{i_k(t)}{Q_k}  = \ell_k(t),
\end{align}
with $Q_k$ being the capacitances and $\ell_k$ the normalised charging rates for cell $k$. The QR-condition for uniform packs can be derived by noting that initially, $ \Delta\bar{v}_{k1}(0) = 0$, and if all cells charge at the same rate then $ \Delta\bar{v}_{k1}(t) = 0$. The problem can then be cast as finding conditions such that $\ell_k(t) = \ell_j(t)$ for all $k$ and $j$. This is determined by rewriting the Kirchhoff voltage laws of Eqn. \eqref{Kirchhoff_F} as 
\begin{align}
0 & = r_1Q_1\frac{i_1(t)}{Q_1}-r_kQ_k\frac{i_k(t)}{Q_k} =r_1Q_1\ell_1(t)-r_kQ_k \ell_k(t)
\end{align}
which implies a uniform pack is achieved with QR matching, as in
\begin{align}
    Q_1r_1 = Q_kr_k.
\end{align}

The following generalises these conditions for uniform packs to the case when there are interconnection resistances. In this case, it is possible to again set $ \Delta\bar{v}_{k1}(t) = 0$ and rewrite the voltage laws of Eqn. \eqref{kv_1_new} as
\begin{subequations}
\begin{align}
&r_{n-1}Q_{n-1}\frac{i_{n-1}(t)}{Q_{n-1}} - (R_n + r_n)Q_n\frac{i_n(t)}{Q_n} = 0,
\\
&r_{n-2}Q_{n-2}\frac{i_{n-2}(t)}{Q_{n-2}} - (R_{n-1} + r_{n-1})Q_{n-1}\frac{i_{n-1}(t)}{Q_{n-1}} \notag \\
&\hspace{4em} -\, R_{n-1}Q_n\frac{i_n(t)}{Q_n} = 0,
\\
&~\, \vdots \notag
\\
&r_1 Q_1 \frac{i_1(t)}{Q_1} - r_2 Q_2 \frac{i_2(t)}{Q_2} - R_2 \sum_{k=2}^n Q_k \frac{i_k(t)}{Q_k} = 0,
\end{align}
\end{subequations}

The pack then behaves uniformly if the following generalised QR-condition is achieved
\begin{subequations}\begin{align}
r_{n-1}Q_{n-1} & = (R_n+r_n)Q_n
\\
r_{n-2}Q_{n-2}  & = (R_{n-1}+r_{n-1})Q_{n-1}+R_{n-1}Q_n, \\ 
& ~\, \vdots \nonumber
\\
r_{1}Q_{1}  & = (r_{2}+R_2)Q_2+R_{2}\sum^n_{k =2}Q_k,
\end{align}\end{subequations}
as in,
\begin{align}\label{rQ}
    r_jQ_j = r_{j+1}Q_{j+1} + R_{j+1}\sum^n_{k = j+1}Q_k.
\end{align}
When all cells have the same capacitances, as in $Q_j = Q_k$ for all $j,k$, then the condition for uniform branch currents in \eqref{rQ}   simplifies to 
\begin{align}\label{R_r}
    r_j = r_{j+1} + R_{j+1}(n-j).
\end{align}
This condition shows that if the cells in packs with interconnection resistances are to receive the same current, then the series resistance of each cell should decrease at a rate proportional to the following interconnection resistance multiplied by its location relative to the back of the pack. Designing the resistances of the pack to satisfy \eqref{R_r} could be achieved by adding additional shunt resistances along with the cells, with the number of added resistors increasing linearly towards the front of the pack.  It is noted that \cite{bhaskar2024heterogeneity} derived an equivalent condition for the case of two cells in parallel and provided experimental data to support the theoretical predictions; condition \eqref{R_r} basically extends that result to the case of $n$ cells in parallel. Finally, it is noted that the expression \eqref{R_r} for uniform pack currents was achieved by working with the analytical solutions of Kirchhoff's laws explored in this paper; such expressions would be challenging to obtain with the numerical solutions. 
%So what is the point of this? Well, the solution of Section \ref{sec:curr_R} is the first solution which quantifies how these current distribute across the pack without having to numerically invert the $A_{22}$ matrix of Eqn. \eqref{A22_2}. It is this insight that we want to explore here, with the solution highlighting the intricacies for how these currents evolve. 

%\color{red} Stress that the analytical solution of the branch currents allowed to compute the Rint as in 4.5. This wouldn't have been possible with the numerical solutions only. In the simultaionts, if we manage to show a speed up with the analytical case for a high number of cells, stress this advantage as well. Add a remark explaining that through potentionmeters, one could balance out the interconnection resistances that are diffeeent due to the manufacturing process of the pack. Moreover, tell that, the potentiometers could be used in closed-loop over the lifespan of the pack to balance also taking into account the different aging of the cells.\color{black}

Figure \ref{fig_all_even1} shows the impact of tuning the resistances across the pack following Eqn. \eqref{R_r}. For the simulations, a pack of 10 LFP cells were connected in parallel using the parameters defined in Section \ref{Simu_res} and charged at 1C. Figure \ref{fig:uneven} shows the evolution of the current distribution when each cell in the pack has the same resistance (as in $r_j = r_k$ for all $j,k = 1,\,2\,,\cdots,\, n$) whereas Figure \ref{fig:even} shows the flat current distribution obtained when the resistances satisfy  \eqref{R_r}.  Exploring this notion further, Fig. \ref{eta} shows the increase in the initial current distribution across the pack as the ratio of the interconnection resistance, $R_k$, to series resistance, $r_k$, increases. In other words,  as the interconnection resistance increases, the variation in the currents also increases as it becomes increasingly challenging for the currents to reach the cells at the back of the pack. 

\begin{figure}[t]
\centering
\includegraphics[width=1\columnwidth]{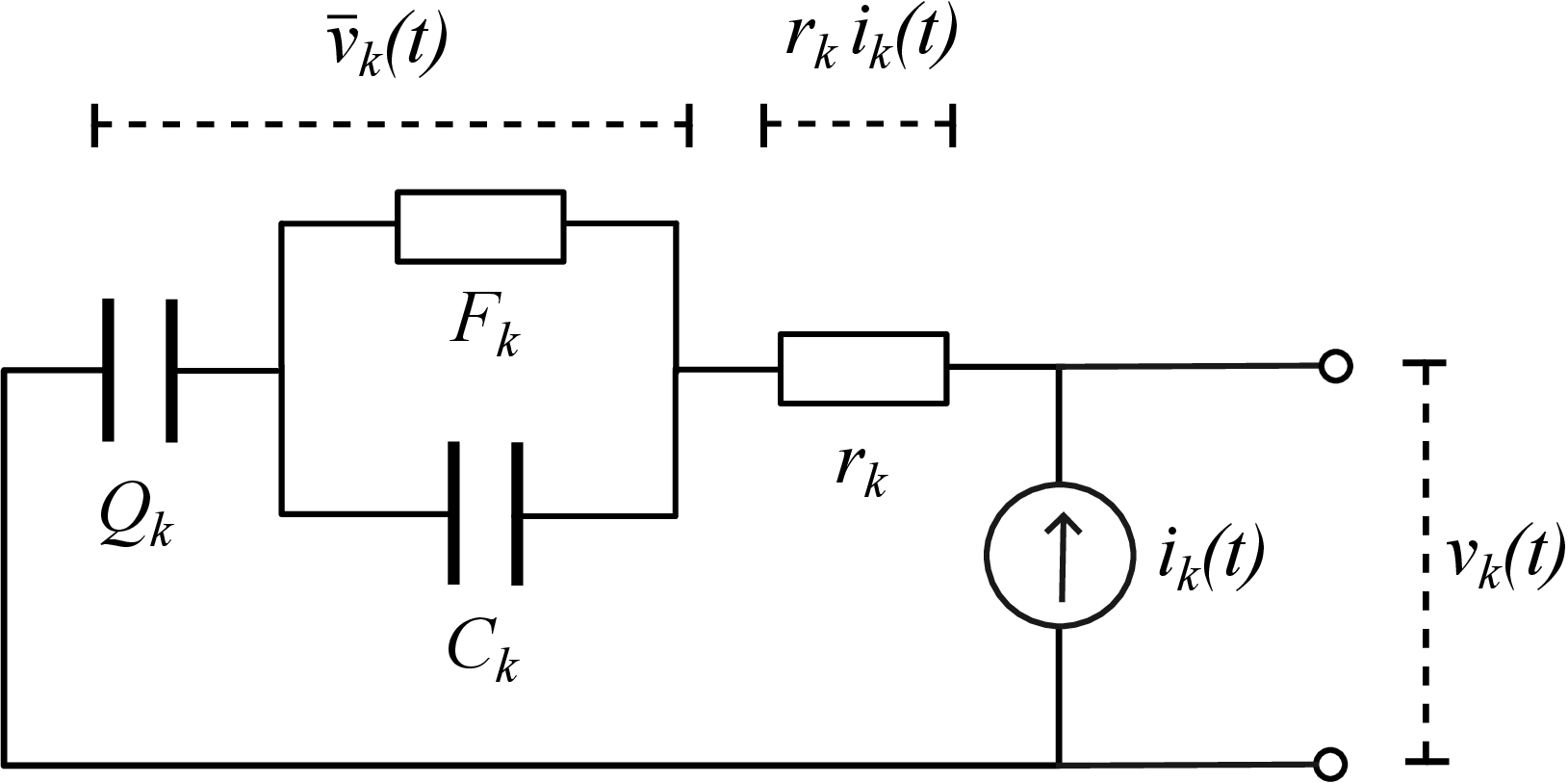}
\caption{Equivalent circuit model of the $k^{\text{th}}$ cell in a parallel-connected pack.}
\label{circuit_fig}
\end{figure}

\begin{figure}[t]
    \centering
    \subfloat[\small SoC–OCV curve for \emph{LG 21700 M50T} cells.]{
        \includegraphics[width=0.8\columnwidth]{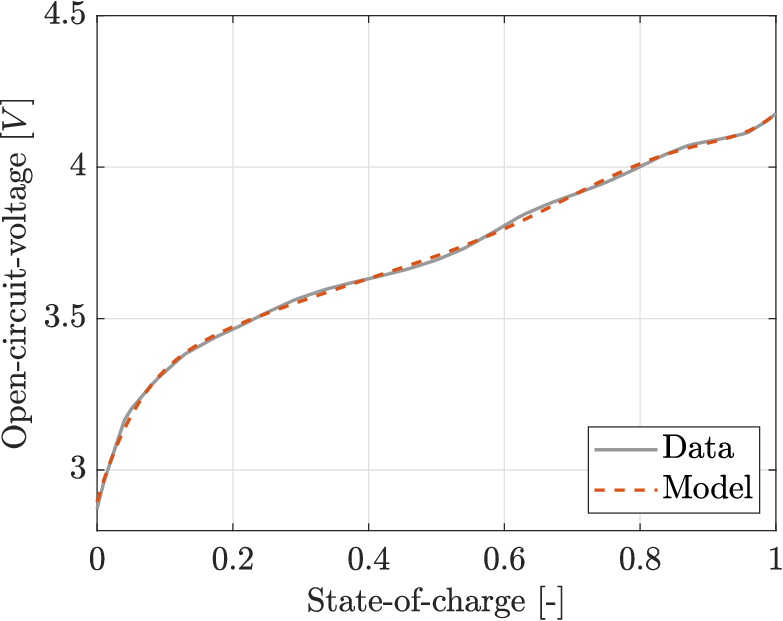}
        \label{SoCOCV_simone}
    }\\%[-0.5ex]
    \subfloat[\small SoC–OCV curve for \emph{K2 LFP26650P} cells.]{
        \includegraphics[width=0.8\columnwidth]{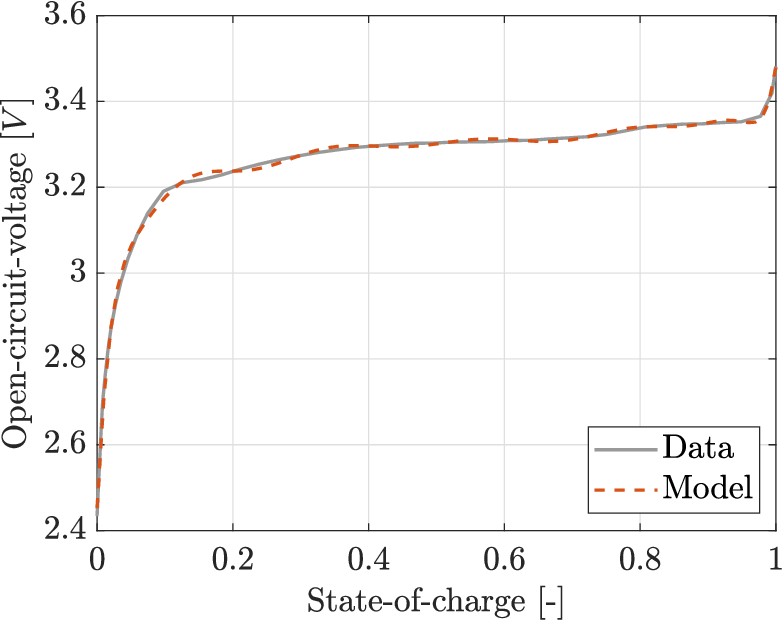}
        \label{SoCOCV_LFP}
    }
    \caption{Open-circuit voltage (OCV) curves used in the model simulations for different cell chemistries.}
    \label{fig:ex1_fb}
\end{figure}

\begin{table}[t]
    \centering
    \caption{Cell-level ECM parameters for model validation using experimental data from four \emph{LG 21700 M50T} cells connected in parallel.}
    \label{table:ECM_LGM50T_param} 
    \begin{tabular}{>{\raggedright\arraybackslash}p{6.8cm} c}
        \hline
        \textbf{Parameter} & \textbf{Units} \\
        \hline
        $r_k(z_k) = -0.056 z_k^3 + 0.116 z_k^2 - 0.073 z_k + 0.0393$ & [$\Omega$] \\
        $F_k(z_k) = -0.02248 z_k^2 - 0.01228 z_k + 0.02551$ & [$\Omega$] \\
        $C_k = 2913.1$ & [$F$] \\
        $Q_k = 4.952$ & [$Ah$] \\
        $OCV(z_k) = 96.7822 z_k^7 - 349.5041 z_k^6 + 512.5251 z_k^5$ & \\
        $\quad\quad -397.1122 z_k^4 + 177.8325 z_k^3 - 46.8445 z_k^2$ & \\
        $\quad\quad + 7.6026 z_k + 2.8955$ & [$V$] \\
        \hline
    \end{tabular}
\end{table}

\begin{table}[t]
    \caption{MSE between simulated and measured cell currents for $R_k = 1$ m$\Omega$ and $R_k = 3$ m$\Omega$.}
    \label{Tab:MSE}
    \centering
    \small
    \begin{tabular}{p{1.3cm} | p{1.3cm} p{1.3cm} p{1.3cm} p{1.3cm}}
        \hline\hline
        \centering $R_k$ & \centering Cell 1 & \centering Cell 2 & \centering Cell 3 & \centering Cell 4 \tabularnewline
        \hline
        1 m$\Omega$ & 0.0105 & 0.0032 & 0.0061 & 0.0298 \\
        3 m$\Omega$ & 0.0123 & 0.0187 & 0.0070 & 0.0141 \\
        \hline\hline
    \end{tabular}
\end{table}

\begin{figure*}[t]
    \centering
    \includegraphics[width=\textwidth]{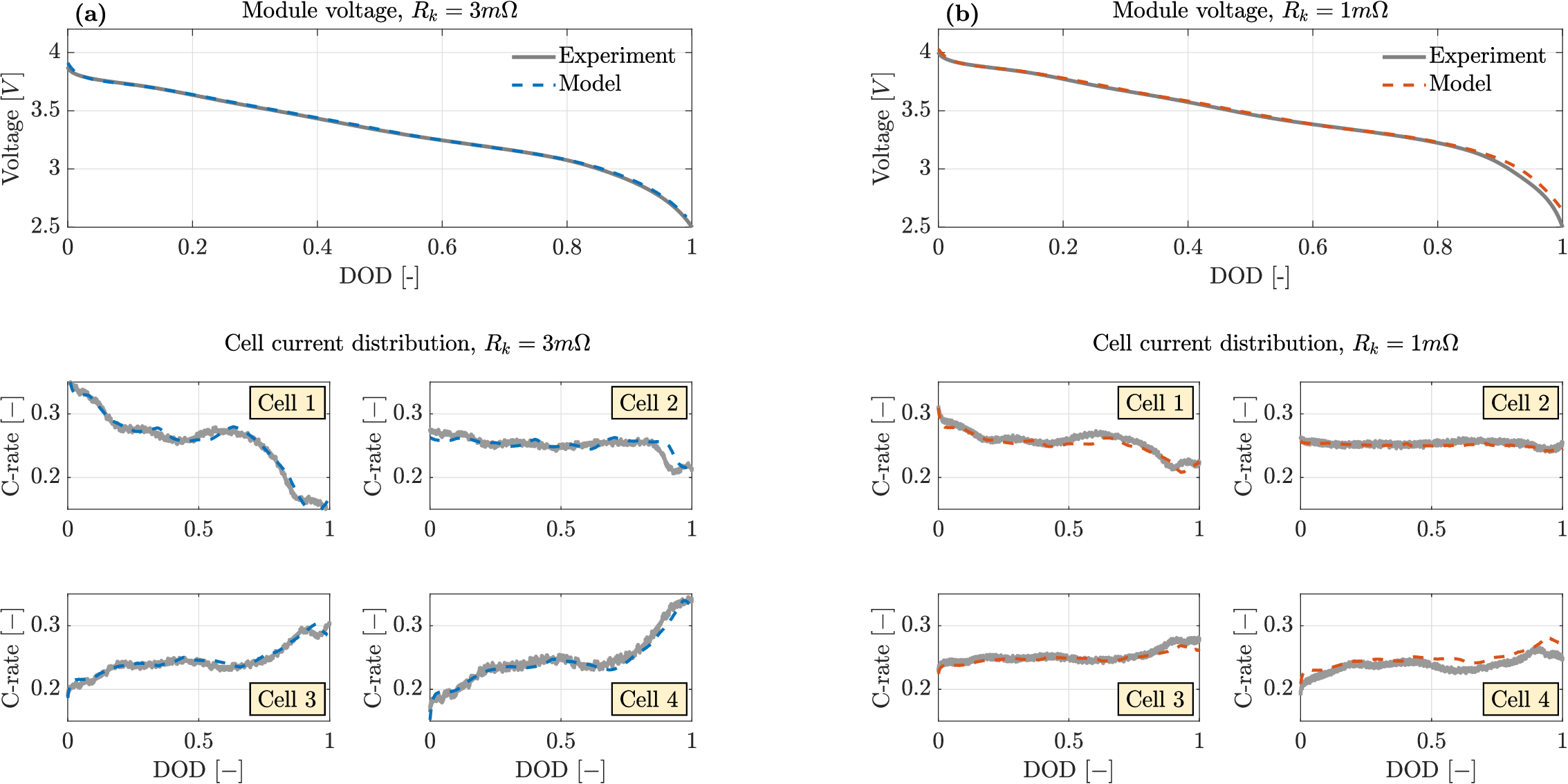}
    \caption{Experimental validation of the parallel pack model using four \emph{LG 21700 M50T} cells connected in parallel. Simulated and measured current and voltage profiles are shown for interconnection resistances of $R_k = 1$ m$\Omega$ and $R_k = 3$ m$\Omega$. DOD denotes the depth-of-discharge of the pack.}
    \label{Fig:validation_plot}
\end{figure*}

\section{Simulation Results and Discussion} \label{Simu_res}
In this section, the proposed modelling approach for parallel connected packs is analysed for two distinct Li-ion chemistries: NMC and LFP.

\subsection{Circuit model for the cells} \label{Sec:ECM}
 For the model simulations, the cell-level dynamics of Eqn. \eqref{cell_dyns} are described by the equivalent circuit models (ECM) of Fig. \ref{circuit_fig}. For cells $k = 1, \dots, n$ (with $n$ being the {number of cells in parallel} in the pack), this circuit maps the current $i_k(t)$ to the voltage $v_k(t)$ by
 \begin{subequations} \label{circuit}
\begin{align}
\frac{d}{dt}
\begin{bmatrix}
z_k(t) \\ w_k(t)
\end{bmatrix}
&=
\begin{bmatrix}
0 & 0 \\
0 & -\frac{1}{F_k C_k}
\end{bmatrix}
\begin{bmatrix}
z_k(t) \\ w_k(t)
\end{bmatrix}
+
\begin{bmatrix}
\frac{1}{Q_k} \\
\frac{1}{C_k}
\end{bmatrix}
i_k(t), \\
v_k(t) &= w_k(t) + \text{OCV}(z_k(t)) + r_k i_k(t),
\end{align}
\end{subequations}
\noindent with $w_k(t)$ being the {relaxation voltages} caused by ion diffusion into active particles,  $z_k(t)$  the state-of-charges (SoC) and $\text{OCV}(z_k(t))$  the open circuit voltages which are nonlinear functions of the SoC.
The {current} going into cell $k$ is denoted $i_k(t)$ and the {voltage} is $v_k(t)$. The voltage contribution from the state variables is 
\begin{align}
\bar{v}_k(t) = w_k(t) + \text{OCV}(z_k(t)).
\end{align} 
 In terms of parameters, $Q_k(t)$ is the {cell's capacitance}, $F_k$ denotes the {resistance} of the RC-pair of the circuit model from Fig. \ref{circuit_fig} and $C_k(t)$ denotes its {capacitance}. It is recognised that the notation $F_k$ for resistance is non-standard for  ECMs but is adopted here since $r_k$ was used for the series resistance and $R_k$ for the interconnection resistance.

By  defining cell $k$'s state as $x_k(t) = [z_k(t), \,w_k(t)]^\top$, the dynamics and voltage from Eqn. \eqref{circuit} can be represented as
\begin{subequations}\label{cell_dyns}\begin{align}
\dot{x}_k(t)  & = \bar{A}_kx_k(t) + \bar{B}_k i_k(t), \quad k = 1, \dots, n,\\
v_k(t)  & = \bar{v}_k(t)+ r_ki_k(t).
\end{align}\end{subequations}
As such, the model is in the standard form of Eqn. \eqref{cell_mod} and so the solutions for the current distributions from Sections \ref{sec:A22inv} and \ref{sec:reccurence} can be applied.

\subsection{Experimental validation}
% [SF comment: I added this section here. Please feel free to change its location or edit it as needed.]

The parallel-connected battery pack model was first validated against experimental data. Four \textit{LG 21700 M50T} cells were connected in parallel with calibrated busbar interconnection resistances of 1 and 3 m$\Omega$, both tested at 25$^{\circ} C$.  Details of the experimental procedures can be found in \cite{piombo2024full} and are not repeated here for the sake of brevity. The tested \textit{LG 21700 M50T} cells were composed of  LiNi$_{0.8}$Mn$_{0.1}$Co$_{0.1}$O$_2$ (NMC 811) cathodes and Graphite-SiO$_x$ anodes and had capacitances of 4.85 Ah.  The complete list of identified parameters for the cells is shown in Table~\ref{table:ECM_LGM50T_param}. The open-circuit voltage, OCV$(z_k)$, of Table \ref{table:ECM_LGM50T_param} was determined by fitting the experimentally measured values, as depicted in Figure~\ref{SoCOCV_simone}, with a polynomial function of $z_k$. Following the approach of \cite{plett2015battery}, the parameters of the dynamic components of the ECM (i.e., $r_k$, $F_k$, and $C_k$) were identified by minimizing the root mean square error between the model voltage and the measured cell voltage during time-varying Hybrid pulse power characterization  (HPPC) current tests. To enhance model accuracy, $r_k$ and $F_k$ were also expressed as polynomial functions of $z_k$, as given in Table \ref{table:ECM_LGM50T_param}. It is emphasised that this state-dependent characterisation of the resistance does not influence the solutions of Section \ref{Resolving} and \ref{sec:slide}-- the only requirement is that the Ohmic drop is linear in the resistance.

% \rdd{The pack model was then validated against experimental data of a 4-cell  pack with calibrated busbar interconnection resistances of 1 and 3 m$\Omega$, both tested at 25$^{\circ} C$. For the pack model, four single-cell ECMs were combined and the branch currents of each cell were calculated using Eqns. \eqref{kvl1} and \eqref{eqn:sn}. 

Figure~\ref{Fig:validation_plot}\,(a)-(b) compares the measured and simulated voltages for $R_k = 1, \text{ and } 3 \text{ m}\Omega$, respectively, showing satisfactory model accuracy.
The accuracy of the estimated cell current was assessed by computing the mean square error (MSE) between the model-estimated cell current and that measured by Hall sensors. The MSE values for each cell, in both module configurations, are reported in Table~\ref{Tab:MSE}. 
During a complete CC discharge cycle at a 0.75C rate, the error remained below 29.8 mA for all 8 cells tested. This corresponded to approximately 0.82\% of the reference cell current at 0.75C (i.e., $0.75 \times 4.85 $ Ah).

\begin{figure*}[t]
    \centering
        \subfloat[\small Current evolution \textit{without} interconnection resistances.]{
        \includegraphics[width=0.23\textwidth]{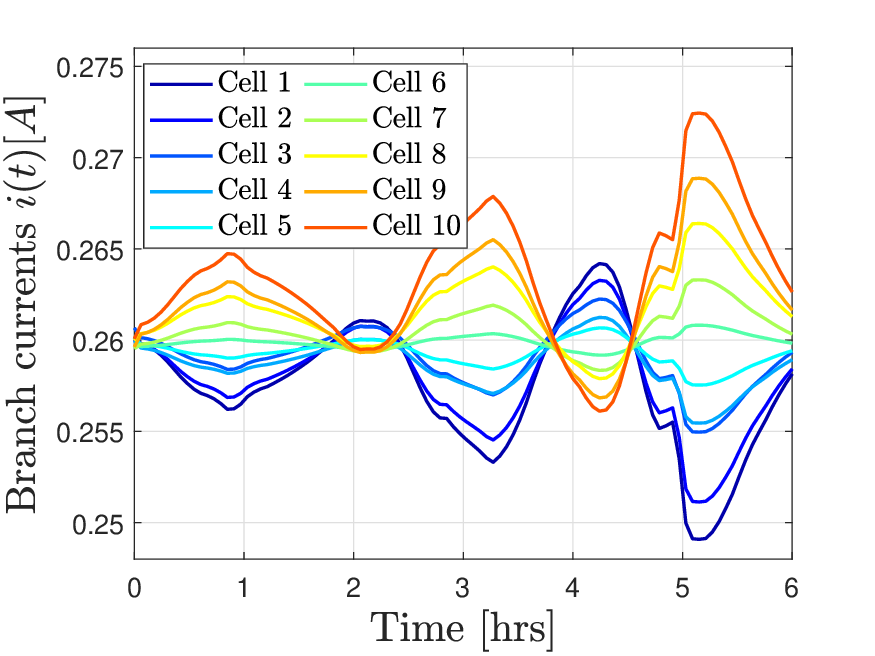}
        \label{BRC_LFP_RZ}
    }
    \hfill
   \subfloat[\small SoC distribution \textit{without} interconnection resistances.]{
        \includegraphics[width=0.23\textwidth]{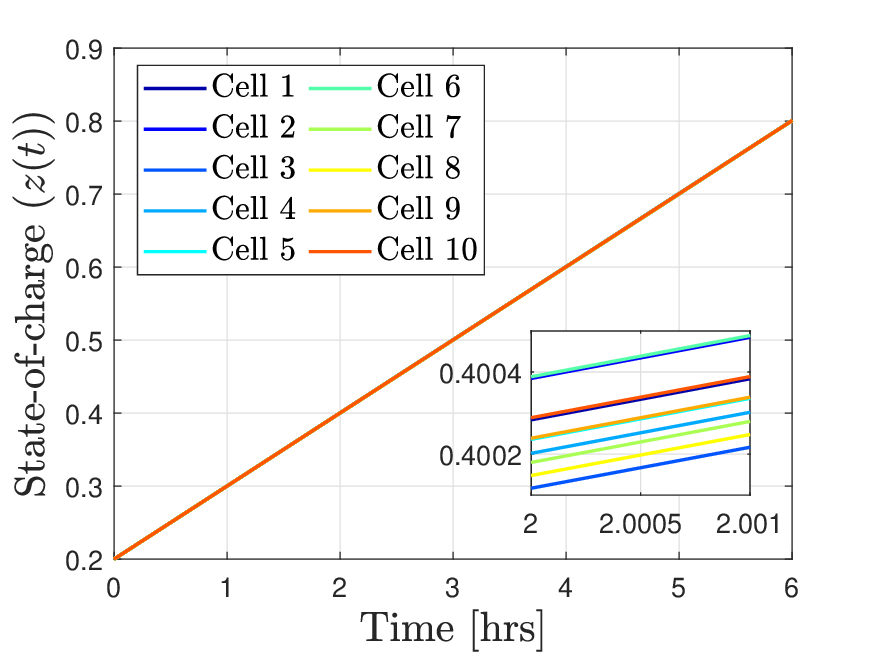}
        \label{SoC_LFP_RZ}
    }
    \hfill
      \subfloat[\small Current evolution \textit{with} interconnection resistances.]{
        \includegraphics[width=0.23\textwidth]{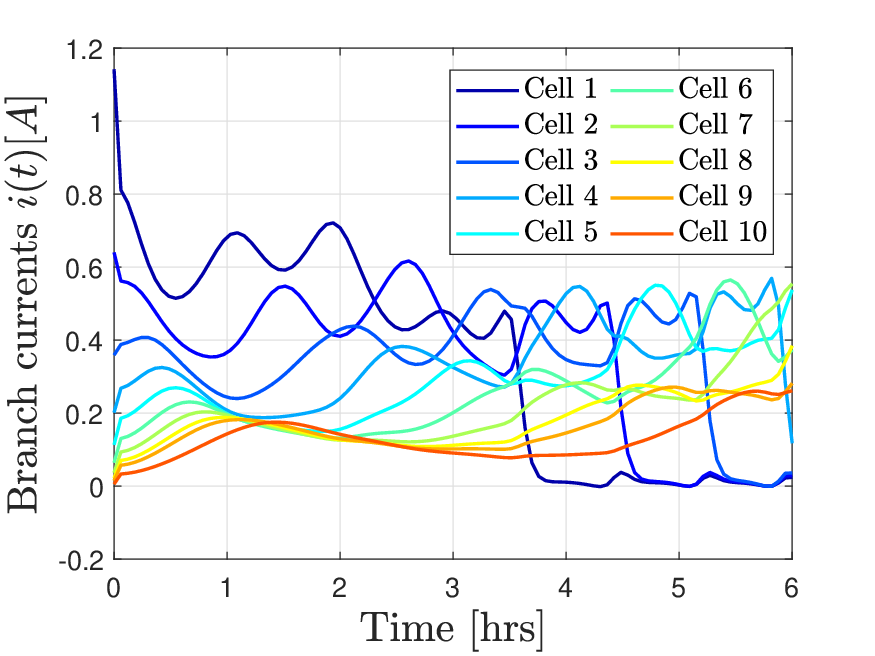}
        \label{BRC_LFP_RNZ}
    }
     \hfill
      \subfloat[\small SoC distribution \textit{with} interconnection resistances.]{
        \includegraphics[width=0.23\textwidth]{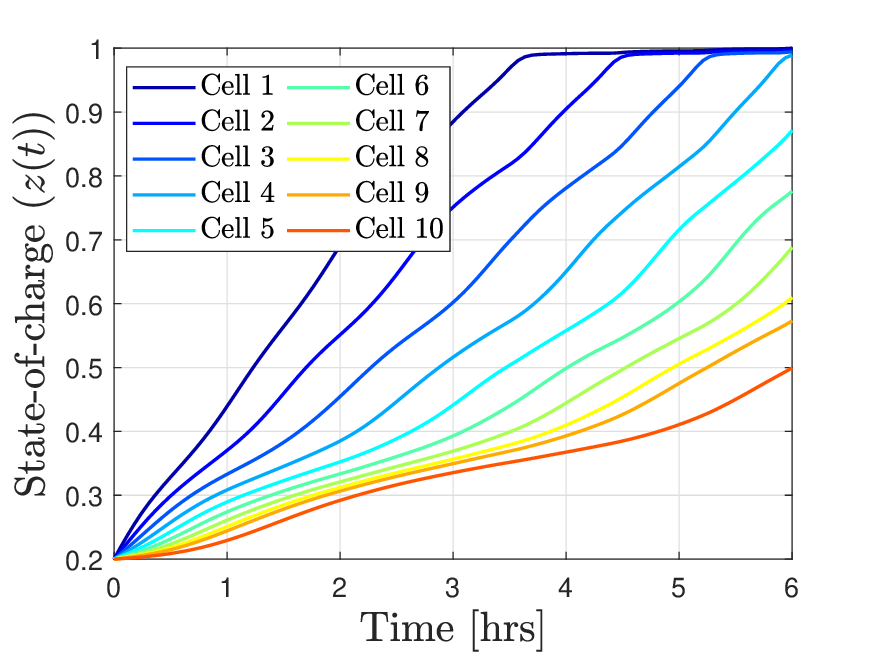}
        \label{SoC_LFP_RNZ}
    }
    \caption{Simulation results for a parallel-connected pack with LFP cells. Subfigures (a) and (b) correspond to the case without interconnection resistances, showing ideal branch currents and SoC distributions. Subfigures (c) and (d) show the impact of including interconnection resistances, where both branch currents and SoC deviate due to imbalance.}
    \label{fig:ex1_fb1}
\end{figure*}

% \begin{figure*}[t]
%     \centering
%     \subfloat[Branch currents of LFP cells \textit{without} interconnection resistances.]{
%         \includegraphics[width=0.45\textwidth]{./Figures/Br_new_LFP_RZ.eps}
%         \label{BRC_LFP_RZ}
%     }
%     \hfill
%     \subfloat[State-of-charge distribution \textit{without} interconnection resistances.]{
%         \includegraphics[width=0.45\textwidth]{./Figures/SOC_Rzero_LFP.eps}
%         \label{SoC_LFP_RZ}
%     }\\[-1ex]
%     \subfloat[Branch currents of LFP cells \textit{with} interconnection resistances.]{
%         \includegraphics[width=0.45\textwidth]{./Figures/Br_new_LFP_RNZ.eps}
%         \label{BRC_LFP_RNZ}
%     }
%     \hfill
%     \subfloat[State-of-charge distribution \textit{with} interconnection resistances.]{
%         \includegraphics[width=0.45\textwidth]{./Figures/SOC_LFP_R_notzero.eps}
%         \label{SoC_LFP_RNZ}
%     }
%     \caption{Simulation results for a parallel-connected pack with LFP cells. Subfigures (a) and (b) correspond to the case without interconnection resistances, showing ideal branch currents and SoC distributions. Subfigures (c) and (d) show the impact of including interconnection resistances, where both branch currents and SoC deviate due to imbalance.}
%     \label{fig:ex1_fb}
% \end{figure*}

\begin{figure}[t]
\centering
\includegraphics[width=0.75\columnwidth]{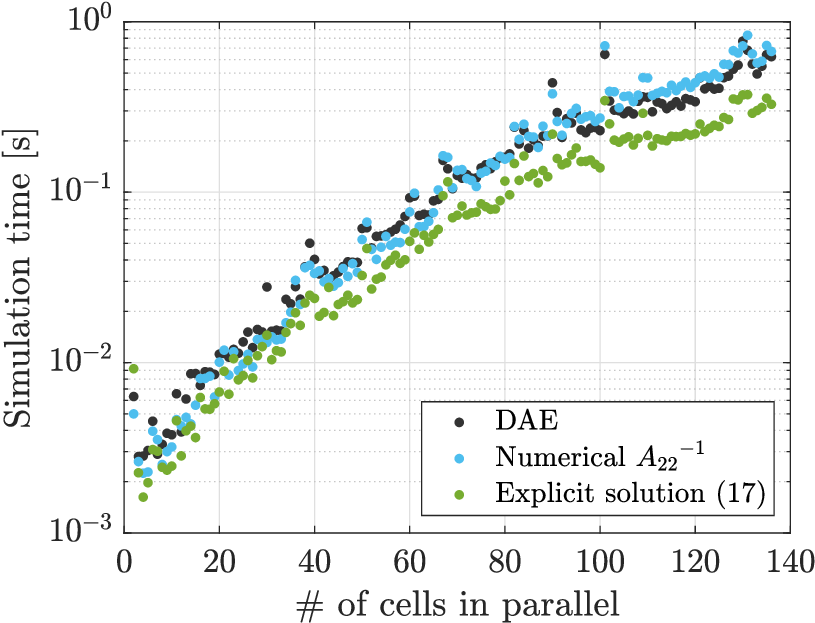}
\caption{Computation times for simulating LFP parallel connected packs models. A comparison between directly solving the DAEs and the formulation in terms of ODEs is shown. The DAEs and ODEs were solved using the \texttt{MATLAB} \texttt{ode15s} function.  }
\label{fig:times}
\end{figure}

\subsection{Parallel connected pack of  LFP cells}
With the parallel pack model validated against experimental data, further simulations were developed to analyse its behaviour. A pack of ten \emph{K2 LFP26650P} cells with LFP cathodes and nominal capacitances of $2.6\,Ah$ was considered for this analysis \cite{tran2021comparative}. For $k = 1, 2, \ldots, 10$, the capacitances were  $C_k = 634\,F$ and the resistances were modelled as Gaussian random variables, with $F_k \sim \mathcal{N}(0.0394, 0.001)\, \Omega$ and $r_k \sim \mathcal{N}(0.0291, 0.001)\,\Omega$. Experimental data of the SoC-OCV and its fit are shown in Fig. \ref{SoCOCV_LFP} \cite{tran2021comparative}.
% \begin{figure}[]
% \centering
% \includegraphics[width=0.8\textwidth]{OCV_SoC_LFP.eps}
% \caption{SoC-OCV curve and fitting for \emph{K2 LFP26650P} cell.}
% \label{SoCOCV_LFP}
% \end{figure}

The evolution of the branch currents without interconnection resistances is shown in Fig. \ref{BRC_LFP_RZ}. The branch currents were relatively uniform, with each cell roughly charging at roughly the same rate (as shown in the SoC curves of  \ref{SoC_LFP_RZ}). However, when interconnection resistances were introduced, the current distribution became uneven, as depicted in Fig. \ref{BRC_LFP_RNZ}.  As a result, the current imbalance caused by the interconnection resistances led to variations in the SoCs, as shown in Fig. \ref{SoC_LFP_RNZ}. These variations could lead to problems such as cell-to-cell variations in state-of-health, localised lithium plating, and thermal gradients in the pack. Recent experimental data such as \cite{bhaskar2024heterogeneity} have demonstrated these cell-to-cell variations can appear in practice, with the current distributions generating inhomogeneous cell heating and, eventually, to the cells degrading at different rates \cite{shi2016effects}. By understanding how the different cells in the pack react through solutions such as Eqn. \eqref{eqn:sn}, improved mechanisms to reduce these pack-level heterogeneities could be designed. 
% This uneven distribution can result in individual cells carrying higher currents, potentially leading to localized heating and accelerated degradation. %Therefore, accounting for interconnection resistances is crucial for ensuring the balanced and reliable operation of the LFP battery pack.

% \begin{figure}[htbp]
%     \centering
%     \begin{minipage}[b]{0.48\textwidth}
%         \centering
%        \includegraphics[width=\textwidth]{Br_new_LFP_RZ.eps}
% \caption{Branch currents of LFP cells in the absence of interconnection resistances.}
% \label{BRC_LFP_RZ}
%     \end{minipage}
%     \hfill
%     \begin{minipage}[b]{0.48\textwidth}
%         \centering
%         \includegraphics[width=\textwidth]{Br_new_LFP_RNZ.eps}
% \caption{Branch currents of LFP cells in the presence of interconnection resistances.}
% \label{BRC_LFP_RNZ}
%     \end{minipage}
% \end{figure}

Finally, Figure \ref{fig:times} shows the computational benefits of simulating parallel connected battery packs as ODEs rather than  DAEs. The figure shows the growth in computation times for simulating LFP cells connected in parallel packs with $I(t) = 2$ C over  1080 s as the pack size increased from $n = 2$ to $n= 135$. For these simulations, the standard deviation in the model parameters was reduced to $10^{-4}$ and $R_k = 10^{-5}$ to ensure the heterogeneities in the states of the large packs did not grow too large and cause additional numerical issues for the solvers.  For each value of $n$, five simulations were run, and Figure \ref{fig:times} shows the average time of those five simulations. The DAE approach involved solving both Eqn. \eqref{cell_mod} and \eqref{kv_1_new} using a ``\emph{mass}" matrix whereas the ODE approach either used the current distributions of Eqn. \eqref{kvl1} or numerically inverted the $A_{22}$ matrix of \eqref{A22_2}. Both the ODEs and DAEs were simulated using the \texttt{MATLAB} \texttt{ode15s} solver. In general, the simulations with the analytical solutions of \eqref{kvl1} were faster than inverting the $A_{22}$ matrix of \eqref{A22_2} or directly solving the DAEs. These gains in simulation times accelerated as the pack size increased; at the end when $n = 135$, the ODE approach of Eqn. \eqref{kvl1} took on average 0.355 s whereas the direct DAE approach with the mass matrix took 0.641 s. The accelerated simulation times  indicate the analytical current distributions could  enable cell-level modelling of the large packs now being deployed in the field, without resorting to lumped approximations. 

% There was a significant increase in simulation time with the DAE approach compared to the ODEs, especially when the pack sizes grew above $n > 30$, highlighting the limitations of directly working with the DAEs when simulating large packs. The difference between the ODE methods, as in between numerically inverting the $A_{22}$ matrix of \eqref{A22_2} and evaluating \eqref{kvl1}, was relatively small, although the explicit solution was generally faster. 

%%
\section*{Conclusions} \label{Conc}
This work introduces explicit solutions for the algebraic equations of Kirchhoff’s laws that characterise the current distributions across parallel connected lithium-ion battery packs. The solutions were expressed in terms of the pack models' states, parameters, and the applied current and allows the model differential-algebraic-equations to be expressed explicitly in terms of ordinary differential equations. The main result was a solution for the current distribution across parallel connected packs with interconnection resistances-- previously only numerical solutions that inverted matrices or applied approximate solutions were used. The results were validated against experimental data and in simulation for packs composed of both NMC and LFP cells and showed, amongst other features, the formation of current fluctuations and distributions across the pack. Compared to the experimental data, the model achieved an accuracy of 0.82\% for the mean squared error of  the branch currents, highlighting its effectiveness. These results could provide deeper insight into the complex dynamics of parallel connections, support pack design to reduce degradation heterogeneity, and enable  pack-level battery management algorithms with robustness guarantees.

\bibliographystyle{IEEEtran}
\bibliography{reff}

\end{document}